\documentclass[fleqn,usenatbib,twocolumn]{mnras}

\usepackage{newtxtext,newtxmath}

\usepackage{xspace}
\usepackage[T1]{fontenc}
\usepackage{ulem}

\usepackage{CJKutf8}

\DeclareRobustCommand{\VAN}[3]{#2}
\let\VANthebibliography\thebibliography
\def\thebibliography{\DeclareRobustCommand{\VAN}[3]{##3}\VANthebibliography}

\usepackage{graphicx}	
\usepackage{amsmath}	

\title[SN Ibn Population synthesis]{Population Synthesis Study on the Binary Origin of Type Ibn Supernovae}

 \author[Ko et al.]{
 Takatoshi Ko$^{1,2}$\thanks{Contact e-mail: \href{mailto:ko-takatoshi@g.ecc.u-tokyo.ac.jp}{ko-takatoshi@g.ecc.u-tokyo.ac.jp}},
 Tomoya Kinugawa$^{3,4,1}$\thanks{Contact e-mail: \href{mailto:kinugawa@shinshu-u.ac.jp}{kinugawa@shinshu-u.ac.jp}},
Daichi Tsuna$^{5,1}$,
 Ryosuke Hirai$^{6,7,8}$,
 and Yuki Takei$^{9,1,6}$
 \\
$^{1}$ Research Center for the Early Universe (RESCEU), School of Science, The University of Tokyo, Bunkyo, Tokyo 113-0033, Japan \\
$^{2}$ Department of Astronomy, School of Science, The University of Tokyo, Bunkyo, Tokyo 113-0033, Japan \\
$^{3}$ Faculty of Engineering, Shinshu University, 4-17-1, Wakasato, Nagano-shi, Nagano, 380-8553, Japan \\
$^{4}$Research Center for Advanced Air-mobility Systems, Shinshu University,  4-17-1, Wakasato, Nagano-shi, Nagano, 380-8553, Japan \\
$^{5}$ TAPIR, Mailcode 350-17, California Institute of Technology, Pasadena, CA 91125, USA \\
$^{6}$ Astrophysical Big Bang Laboratory, Pioneering Research Institute, RIKEN, 2-1 Hirosawa, Wako, Saitama 351-0198, Japan\\
$^{7}$School of Physics and Astronomy, Monash University, Clayton, Victoria 3800, Australia
\\
$^{8}$OzGrav: The ARC Centre of Excellence for Gravitational Wave Discovery, Australia\\
$^{9}$ Yukawa Institute for Theoretical Physics, Kyoto University, Kitashirakawa-Oiwake-cho, Sakyo-ku, Kyoto, Kyoto 606-8502, Japan\\
}

\begin{document}
\label{firstpage}
\pagerange{\pageref{firstpage}--\pageref{lastpage}}

\maketitle

 \begin{abstract}
Type Ibn supernovae (SNe) are a class of SN explosions whose progenitors are surrounded by dense helium-rich circumstellar matter (CSM). Some models have been proposed for how to form the dense CSM, with promising scenarios involving either binaries with a low-mass ($\lesssim 3~M_\odot$) helium (He) star, or mergers following common envelope phases between a He star and a compact object. Using rapid binary population synthesis calculations, we estimate the event rate of these channels and compare it with the observed SN Ibn rate. We find that exploding low-mass He stars in close binaries (of separations $\lesssim$ a few 100 $R_\odot$) can be sufficiently produced to account for the observed event rate of SN Ibn, while the merger scenario can likely account for only a fraction of these SNe. We discuss the types of companions expected in the low-mass He star scenario, finding massive main sequence stars ($10$--$20\ M_\odot$) to be typical, with a potentially non-negligible fraction ($<10\%$) of binaries with {white dwarf (WD)} companions that have long delay times of up to $100$ Myrs. 
\end{abstract}




\section{Introduction}
Recent optical surveys have identified numerous interaction-powered supernovae (SNe), which are generally more luminous than typical SNe and display narrow emission lines -- hydrogen in the case of SN IIn~\citep[e.g.,][]{1990MNRAS.244..269S,1997ARA&A..35..309F}, helium for SN Ibn, carbon for SN Icn~\citep[e.g.,][]{2022Natur.601..201G,2022ApJ...938...73P}, and silicon/sulphur for SN Ien \citep[][]{2024arXiv240902054S}. Such interaction-powered SNe are thought to exhibit these characteristics as a result of collisions between the SN ejecta and dense circumstellar medium (CSM) surrounding the progenitor. The dense CSM is expected to form shortly before the explosion, and likely holds key information about the final evolutionary stages of massive stars. Various scenarios have been proposed to explain the presence of dense CSM, including unsteady nuclear burning driven by turbulent convection \citep[e.g.,][]{Smith2014turb}, pulsational instabilities \citep[e.g.,][]{Woosley2002,Woosley2007}, wave-driven mass-loss triggered by nuclear burning in the core \citep[e.g.,][]{Quataert2012,Shiode2014}, and common envelope (CE) evolution involving a binary companion \citep[e.g.,][]{2012ApJ...752L...2C}. However, a unified understanding of the underlying mechanisms remains elusive.

Type Ibn SNe are a unique class of interacting SNe, with signatures of CSM primarily composed of helium. They are relatively rare, with the event rate estimated to be approximately 1--2~\% of all core-collapse SNe (CCSNe)~\citep[][]{2008MNRAS.389..113P,2022ApJ...927...25M,Ma25}. SNe Ibn have initially been proposed to originate from Wolf-Rayet (WR) stars -- the terminal stages of very massive stars with initial masses of $\gtrsim 25~M_\odot$ that have lost their hydrogen-rich envelopes via strong stellar winds, or low mass helium (He) stars whose envelopes have been stripped off by binary interactions. 
Early studies on the prototype SN Ibn SN 2006jc have suggested explosions of massive WR stars, based on the detection of a pre-SN outburst in 2004 reminiscient of luminous blue variables~\citep[e.g.,][]{2007ApJ...657L.105F,2007Natur.447..829P,2008ApJ...687.1208T}. However, recent studies suggest that high-mass WR stars may not be the sole progenitors. For instance, low-mass He stars (of mass $\lesssim$ 3--5$~M_\odot$) that undergo binary interactions have been suggested to explain the spectral observations~\citep[e.g.,][]{2022A&A...658A.130D} and the dense CSM \citep[e.g.,][]{wu2022extreme,2024ApJ...977..254D,2024OJAp....7E..82T,2024arXiv241209893E}. Later observations have found a surviving companion in SN 2006jc~\citep[e.g.,][]{2016ApJ...833..128M,Sun20}, indicating that some SNe Ibn indeed originate from binary systems. 

Additionally, the environments of some SNe Ibn appear to favor binaries as the formation channel. For example, SN 2023tsz was observed in a dwarf galaxy with low metallicity~\citep[][see also \citealt{2023MNRAS.523.2530D}]{2025MNRAS.536.3588W}. In such environments, formation of WR stars due to wind mass loss is difficult, supporting that SNe Ibn may originate from the explosions of low-mass He stars~\citep[e.g.,][]{2016ApJ...833..128M,2022A&A...658A.130D,wu2022extreme,2024ApJ...977..254D,2024arXiv241209893E}. Furthermore, while CCSNe from massive stars are generally found in galaxies with high star formation rates (SFRs)~\citep[e.g.,][]{2015A&A...580A.131T,2021ApJS..255...29S}, some SNe Ibn have been reported in galaxies with low SFR. PS1-12sk is an example of an SN Ibn observed in the outskirts of a low-SFR galaxy, which disfavors a massive star origin for this event~\citep[][see also \citealt{2019ApJ...871L...9H}]{2013ApJ...769...39S}. 

These cases indicate that not all SNe Ibn originate from single very massive stars that evolve into WR stars, and He stars formed in binary systems may potentially be a dominant channel of SNe Ibn. This has previously been suggested for stripped envelope supernovae (SESNe), and binary population modeling have successfully reproduced the event rates and ejecta properties inferred in these SNe \citep[e.g.,][]{1992ApJ...391..246P,2007ApJ...670..747K,2008MNRAS.384.1109E,2010ApJ...725..940Y}. However for SN Ibn, no studies using binary population models have been conducted to test the binary scenario, as well as to constrain the binary evolution channels leading to SN Ibn.

In this study, we perform rapid binary population synthesis calculations to estimate the event rate of SNe Ibn originating from binary systems, based on two proposed scenarios for producing these SNe. We first consider a scenario in which a low-mass He star within a specific mass range undergoes mass-loss shortly before CCSN and explodes as an SN Ibn \citep{wu2022extreme,2024arXiv241209893E}, referred here as the Low-mass He Star Scenario. We then explore another scenario in which a He star merges with a neutron star (NS) or a black hole (BH) and evolves into an SN Ibn \citep{2012ApJ...752L...2C,2022ApJ...932...84M}, referred to as the Merger Scenario. By comparing these estimates with observations, we provide predictions about the progenitor properties and the properties of the companion stars which would be left after the SN. 

This paper is constructed as follows. In Section~\ref{sec:Methods}, we explain the setup of the binary population synthesis calculation, and the method to estimate the SN event rates from the binary models. In Section~\ref{sec:results}, we overview the characteristics of He star binaries generated in the population synthesis code, and compare the SNe Ibn event rates predicted from specific binary models to the observed rates. Then, in Section \ref{sec:discussion} we discuss the progenitor and remnant features of SNe Ibn which may be testable by future observations of these SNe. We conclude in Section \ref{sec:conclusion}.

\section{Methods}
\label{sec:Methods}
We use the rapid binary population synthesis code described in \citet{2014MNRAS.442.2963K,2016MNRAS.456.1093K,2024MNRAS.532.3926K}, which is based on the Binary Stellar Evolution (BSE) code \citep{2000MNRAS.315..543H,2002MNRAS.329..897H}.
This code updates the treatment of mass transfer and CE evolution from those of BSE \citep{2002MNRAS.329..897H}, as described below. 
In this section, we describe the key features of our population synthesis code, and the method for estimating the rates of SNe Ibn and other events from the binary systems generated by the code.

\subsection{Binary population synthesis code}\label{sec:pop_syn}
The initial conditions of the binaries follow Salpeter's initial mass function (IMF) $\propto M^{-2.35}$ \citep{1955ApJ...121..161S} for the initial primary mass $M_{\rm prim}$, a uniform initial mass-ratio distribution from $0.1~M_\odot/M_{\rm prim}$ to $1$ \citep{2007ApJ...670..747K,2012ApJ...756...50K,2022A&A...665A.148S}, a log-flat initial separation function $\propto 1/a$ from $a_{\rm min}$ to $10^6~R_\odot$ \citep{1983ARA&A..21..343A}, and a thermal equilibrium distribution function for the eccentricity $\propto e$ from 0 to 1 \citep{1975MNRAS.173..729H, 1991A&A...248..485D}. Here $a_{\rm min}$ is the minimum separation where the binary cannot interact at zero age main sequence (ZAMS). We fix the metallicity to the Solar value as the host environments of SN Ibn do not show strong differences with respect to normal CCSNe \citep{2021ApJS..255...29S}, although the samples of SNe Ibn are still small to draw a firm conclusion.

In our binary models, we focus on those that become progenitors of CCSNe and therefore restrict $M_{\rm prim}$ to a range $3~M_\odot\leq M_{\rm prim}\leq 100~M_\odot$. The lower limit of $3~M_\odot$ is chosen to produce a large enough sample of binaries of our interest that lead to CCSNe, while ensuring that we capture CCSNe from the lowest-mass progenitors. We take into account this modified lower limit when estimating the SN event rates in Section \ref{sec:rates}.

We randomly generate the binary systems based on the aforementioned initial distribution functions, and evolve them for 15 Gyr. The parameters for the population synthesis models that we calculated are summarized in Table~\ref{tab:init}. The threshold for CCSN is expected to be where the mass of the carbon-oxygen (CO) core is comparable to the Chandrasekhar limit. Here we adopt a threshold of $1.44~M_\odot$ for the CO core mass at core carbon ignition, for the star to undergo a CCSN. {We note that stars whose CO core masses are from $1.38~M_{\odot}$ to $1.44~M_{\odot}$ have been proposed to lead to an electron-capture supernova (ECSN) \citep[e.g.][]{1980PASJ...32..303M,2013ApJ...771...28T} and this mass range may significantly expand if the progenitor experienced binary interactions, as they can suppress the second dredge-up \citep[e.g.,][]{2004ApJ...612.1044P,2010NewAR..54..140V,2021ApJ...920L..37W}. However, since the number of ECSNe from such a narrow mass range is expected to be small compared to the number of other CCSNe considered in this study, contributions of ECSNe are not included in our calculation for simplicity.}

When calculating the change in the binary orbit due to a SN explosion, we assume that the mass is instantaneously ejected \citep{1995MNRAS.274..461B}.
For NS formation, in addition to the spherical mass ejection from the progenitor we consider a natal kick velocity following an isotropic, Maxwell distribution with $\sigma = 250~\mathrm{km\ s}^{-1}$\citep[e.g.][]{2005MNRAS.360..974H}. 
We assume the natal kick for BH remnants is zero for simplicity, although a non-negligible fraction of BHs with larger kicks ($\sim 100\ {\rm km\ s^{-1}}$) have been recently suggested \citep{Nagarajan24,Burrows24}.

For the binary orbital evolution due to mass transfer, we mainly follow the formalism adopted by \cite{2024MNRAS.532.3926K}. 
For Roche-lobe overflow, we define the mass transfer efficiency $\beta$ as
    $\dot{M}_2 = -\beta \dot{M}_1$, 
where $M_1$, $M_2$ are respectively the mass of the donor and the accretor\footnote{Note that they do not necessarily mean the primary ({more massive} one) and secondary ({less massive} one) at ZAMS.}. In this work, we assume for simplicity that the mass transfer efficiency parameter $\beta$ {($=$ 0.5 or 1 as listed in table~\ref{tab:init})} is universal over all systems at any evolution stage, except when the recipient star is a compact object, i.e. NS, BH, or white dwarf (WD). In this case, $\beta$ is adjusted so that the accretion rate onto the compact object is limited by the Eddington mass accretion rate~\citep{2002MNRAS.329..897H,2014MNRAS.442.2963K}. 
When mass is lost from the system during mass transfer, the specific angular momentum carried away by the ejected mass is that of the donor star {if the accretion rate onto the accretor remains below the Eddington limit, for both compact and non-compact object accretors. However, if the accretion rate onto the accretor reaches the Eddington limit, which is generally achieved only for compact accretors,} the lost mass is assumed to carry the specific angular momentum of the accretor.

The binary enters a CE phase when mass transfer becomes unstable, whose criterion is determined based on Section 2.1 of \cite{2024MNRAS.532.3926K}. To model the CE phase of the binary system two parameters $\alpha$ and $\lambda$ are used, which respectively parameterize the efficiency of CE ejection and the donor structure. The separation after the CE phase is calculated by the energy formalism \citep{1984ApJ...277..355W,deKool90} defined by
\begin{equation}
    \alpha \left( \frac{GM_{\mathrm{c,1}} M_2}{2a_{\mathrm{f}}} - \frac{GM_1 M_2}{2a_{\mathrm{i}}} \right) = \frac{GM_1 M_{\mathrm{env,1}}}{\lambda R_1}.
\end{equation} Here, $a_{\mathrm{i}}$ and $a_{\mathrm{f}}$ respectively represent the initial (pre-CE) and final (post-CE) orbital separations, $M_1, M_{\mathrm{c,1}}, M_{\mathrm{env,1}} = M_1 - M_{\mathrm{c,1}}$ are respectively the donor's total mass, core mass and envelope mass, $M_2$ is the accretor's mass, and $R_1$ is the donor's radius before the CE phase. When $a_f$ is less than the sum of the post-CE remnants' radii, we assume the stars have merged. It can be seen that $a_f$ scales with the product $\alpha\lambda$, which we vary in this work also assuming it is universal over all systems. We also assume that if the donor star initiates the CE phase while in the Hertzsprung gap, the system will merge rather than survive the CE phase~\citep[e.g.,][]{2007ApJ...662..504B}.

In our study, we explore six sets of binary parameters: $(\beta, \alpha \lambda) = (0.5, 0.1), (0.5, 1), (0.5, 10), (1, 0.1), (1, 1),$ and $(1, 10)$. We simulate $10^6$ binary systems for each parameter set.

\begin{table}
\centering
\begin{tabular}{|l|l|}
\hline
\textbf{Parameter}               & \textbf{Distribution/Value}            \\ \hline
Initial mass function            & Salpeter                 \\ \hline
Initial mass ratio distribution          & Flat                     \\ \hline
Initial separation distribution  & Log flat                 \\ \hline
Initial eccentricity function    & $\propto e$                       \\ \hline
Natal kick velocity (\(\sigma\)) & 250 km s\(^{-1}\)        \\ \hline
CE parameter (\(\alpha\lambda\)) & (0.1, 1, 10)             \\ \hline
Mass transfer efficiency (\(\beta\)) & (0.5, 1)              \\ \hline
\end{tabular}
\caption{Summary of the initial distributions of the binary parameters and our binary evolution parameters adopted in this work. A total of six combinations for $(\alpha\lambda,\beta)$ are considered, with $10^6$ binaries for each combination.}
\label{tab:init}
\end{table}

\subsection{Estimating the event rate}
\label{sec:rates}
When considering a specific class of events from binary origin, its event rate in our model can be expressed as \citep[e.g.,][]{2024MNRAS.532.3926K}
\begin{equation}
    R_{\rm event} = \frac{N_{\rm binary}}{2N_{\rm binary}+N_{\rm single}}\times \frac{\rm SFR}{\bar{M}} \frac{\int_{3M_\odot}^{100 M_\odot}\rm IMF~dM}{\int_{0.1M_\odot}^{100 M_\odot} \rm IMF~dM}\times \frac{N_{\rm event}}{N_{\rm tot}},
\end{equation}
where SFR is star formation rate, $\bar{M}$ is the averaged initial star mass ($\bar{M} = \int_{0.1M_\odot}^{100 M_\odot}(M\times\mathrm{IMF)~dM}/{\int_{0.1M_\odot}^{100 M_\odot}\rm IMF~dM}$), IMF is the initial mass function, and $N_{\rm event}$ is the number of events obtained by our population synthesis code out of $N_{\rm tot} = 10^6$ binary systems generated. $N_\mathrm{binary}$ and $N_\mathrm{single}$ are the number of binary systems and the number of single stars, respectively. The event rate depends on the SFR and the minimum mass of the IMF adopted in the integral. In this work we only consider the event rate with respect to CCSNe when comparing with observations, and in this case the relative fractions will be independent of these parameters. We consider primary stars with mass ranging from 3 to 100 $M_\odot$, and thus the integration limits in the numerator are set accordingly. We adopt a binary fraction $N_\mathrm{binary}/(N_\mathrm{single} + N_\mathrm{binary})$ of 1/2 \citep[e.g.][]{2013A&A...550A.107S,Tian2018}.

In order to estimate the fraction of events with respect to CCSNe, the event rate of the CCSN is also required. As CCSNe can originate not only from binary systems but also from single massive stars, the CCSN rate is expressed as
\begin{eqnarray}
    R_{\rm CCSN} &=& \frac{1}{3}\times \frac{\rm SFR}{\bar{M}} 
    \frac{\int_{3M_\odot}^{100 M_\odot}\rm IMF dM}{\int_{0.1M_\odot}^{100 M_\odot} \rm IMF dM} \times \frac{N_{\rm CCSN,bin}}{N_{\rm tot}}\nonumber \\
    &+& \frac{1}{3}\times \frac{\rm SFR}{\bar{M}} 
    \frac{\int_{M_{\rm SN, min}}^{100 M_\odot}\rm IMF dM}{\int_{0.1M_\odot}^{100 M_\odot} \rm IMF dM},
\end{eqnarray}
where $M_{\rm SN, min}\approx 8.0~M_\odot$ is the minimum ZAMS mass for a single star to undergo CCSN based on our criterion, which correspond to a CO core mass of $1.44~M_\odot$. The method to extract the CCSNe events from our binary population is summarized in Appendix~\ref{sec:app_CCSN}. We find that the number of CCSNe ($N_{\rm CCSN,bin}\sim 4\times 10^5$) from our binary population is almost independent of the adopted ($\beta,\alpha\lambda$), as summarized in Table~\ref{tab:result}.
 
The fraction with respect to CCSNe of a given class of event from our binary population is calculated as
\begin{eqnarray}\label{eqn:ratio}
   && r_{\rm event, bin} \nonumber\\
    &=& \frac{\int_{3M_\odot}^{100 M_\odot}\mathrm{IMF dM}\times N_{\rm event}}{\int_{3M_\odot}^{100 M_\odot}\mathrm{IMF dM} \times N_{\rm CCSN,bin} 
    + \int_{M_{\rm SN, min}}^{100 M_\odot}\mathrm{IMF} \mathrm{dM} \times N_{\rm tot}} \nonumber \\
    &\approx& \frac{N_{\rm event}}{N_{\rm CCSN,bin} + 0.26N_{\rm tot}}.
\end{eqnarray} 
We specifically consider the fraction of SESNe and Type Ibn SNe with respect to CCSNe, using our binary models that lead to stripped stars explained in the next section. 

Equation (\ref{eqn:ratio}) implicitly assumes that all core-collapse events result in a SN explosion, while we note that some core-collapse events may lead to a quiet BH formation without explosion. However such cases are generally rare \citep[e.g.,][]{Neustadt21}, and a SN-like explosion with a BH remnant can occur even from non-rotating progenitors~\citep[e.g.,][]{Quataert19,Antoni23,Soker24,Burrows24}. Furthermore, when calculating the ratio of specific SNe explosions with respect to CCSNe, such BH-forming cases are included in both categories, so their overall effects on our results are likely small.

\section{Results}\label{sec:results}
\subsection{Features of binary systems containing exploding He stars}

\begin{figure*}
 \centering
 \includegraphics[width=\linewidth]{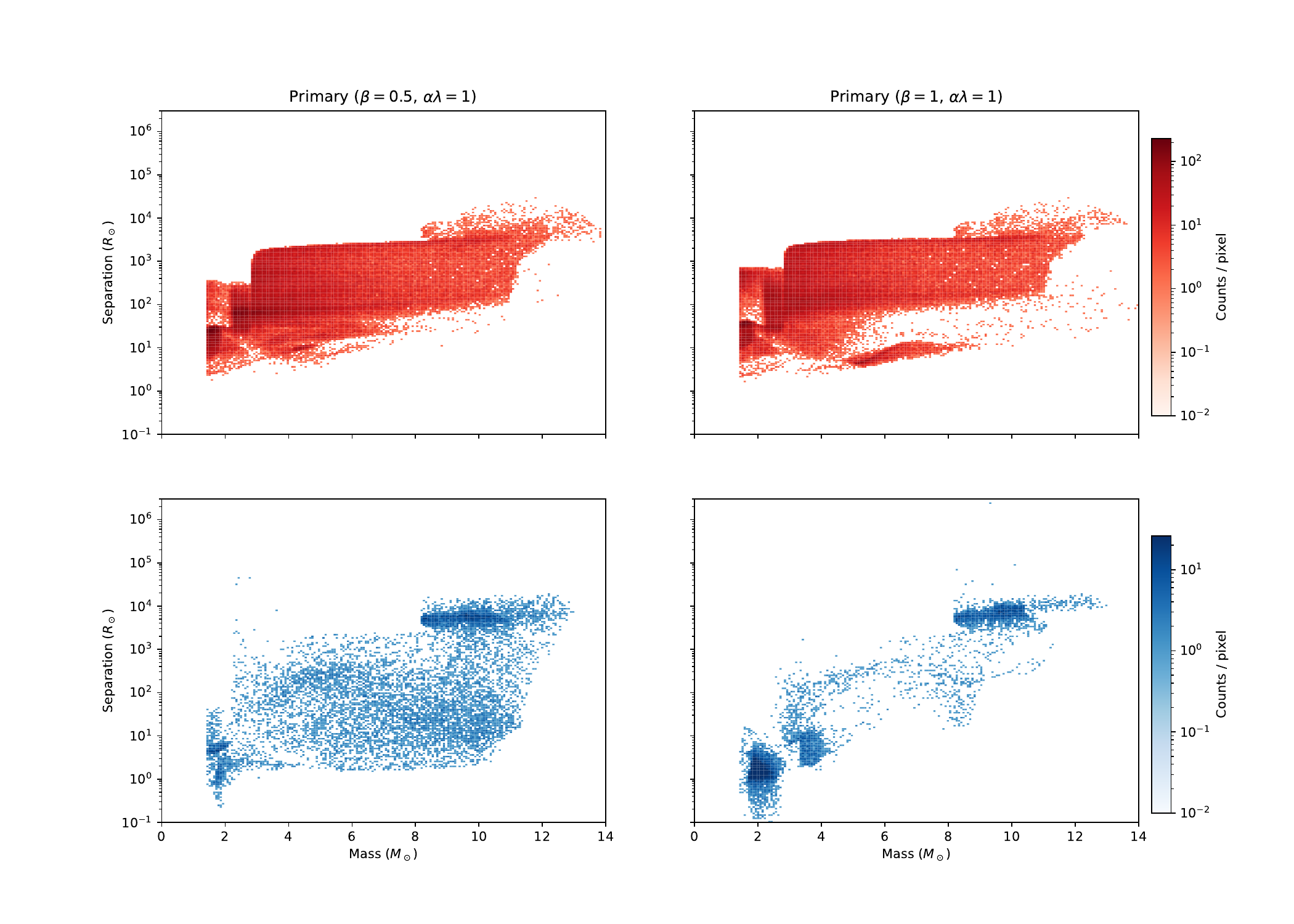}
\caption{
{Counts heat maps of }binary separation versus He star mass of binary systems before core-collapse of the He star, for representative binary parameters of $(\beta, \alpha \lambda) = (0.5,1)$ and $(1,1)$. 
Top panels indicate binary systems where the initially more massive primary is the He star (whose companions are typically stars), while bottom panels represent those where the initially less massive secondary is the He star (whose companions are typically NSs or BHs). {Pixel size along both the $X$ and $Y$ axes is $1/200$ of the plotted range (i.e., a $200\times200$ grid).
}}
 \label{fig:mass_sep}
\end{figure*}

Before considering the detailed channels of SN Ibn, we first outline the overall demographics of He stars just before they undergo explosion in BSE. Specifically, we extract He stars that are just about to begin core carbon burning, excluding those that underwent merger with a H-rich companion (which are still labeled as He stars in the BSE code). The details of this extraction method is explained in Appendix~\ref{sec:app_Ibc}. There are approximately $2-3\times10^5$ binary systems that contain the He stars extracted in this manner.

The pre-core collapse He star mass and binary separation are shown in Figure \ref{fig:mass_sep}, for representative binary parameter sets of $(\beta, \alpha \lambda) = (0.5, 1)$, $(1, 1)$. 
Here, the upper panels in red show systems in which the initially more massive primary is the pre-core collapse He star, whereas the lower panels in blue show systems where the secondary is the He star. The former systems generally have stellar (typically main-sequence) companions, whereas the latter generally have compact object (NS, BH and white dwarf) companions.

The binary systems in the top panels, where the primary is the He star, can be broadly classified into three categories in this mass-separation diagram:  
(1) systems with separations of $\sim 10$--$1000~R_\odot$ that created the He star by stable mass transfer, which constitute the majority of the population;  
(2) systems that have experienced (and survived) a CE phase, resulting in a significant shrinkage of the separation --- these correspond to the cluster of points with He star mass $4-8 \,M_\odot$ and separation of $\sim1$--$10\,R_\odot$; 
(3) systems that have evolved without significant binary interaction, where a massive star becomes a He star solely via stellar winds --- these correspond to the population with He star mass $> 9\,M_\odot$ and separation larger than $\sim2000\,R_\odot$.

The binaries in the bottom panels, representing binary systems of He stars with compact object companions, can also be categorized into three groups in a similar way to the top panels based on their most recent binary interaction. We find that category (2), referring to binaries which experienced CE between the He star progenitor and the compact object, span a parameter space different from that in the top panels. These systems are concentrated in regions with small separations and He star masses of $\lesssim 2\ M_\odot$. For $\beta=0.5$, the binaries in the bottom left panel span a similar distribution as those in the above panel. On the other hand, for $\beta = 1$ the number of tight binaries of intermediate He star masses ($4$--$8~M_\odot$) formed through CE is noticeably reduced. This is partly because the case for $\beta=1$ often results in mass reversal of the binaries via mass transfer, which tends to increase the separation once the accretor becomes more massive. The first mass transfer phase in the system widens the separation, making them more susceptible to binary disruption upon SN when strong natal kicks are involved. As a result, most systems with NS companions end up being disrupted before it can tighten the orbit via CE evolution. This is evidenced by the fact that for systems with He star mass of $4$--$8\,M_\odot$, the companion is found to be biased towards BHs. For the more massive systems where the primary becomes a BH, some close binaries with He star mass around $8\,M_\odot$ remain intact.

\begin{table*}
\centering
\begin{tabular}{|c|c|c|c|c|c|c|c|c|c|c|c|c|}
\hline
\multicolumn{1}{|c|}{} & \multicolumn{3}{c|}{$\beta = 0.5$} & \multicolumn{3}{c|}{$\beta = 1$} \\
\hline
$\alpha\lambda$  &  $10^{-1}$ & $10^{0}$ & $10^{1}$ & $10^{-1}$ & $10^{0}$ & $10^{1}$\\
\hline
$N_{\rm CCSN,bin}$  & 384896 & 411292 & 402078 & 431479 & 440417 & 448646 \\
\hline
$N_{\rm SESN, bin}$ ($r_{\rm SESN, bin}$) & 95452 (0.15)& 119969 (0.18)& 133064 (0.20)& 77864 (0.11)& 105343 (0.15)& 110397 (0.16)\\
$r_{\rm SESN, bin}+r_{\rm SESN,sin}$ & 0.24 &0.27&0.29&0.20&0.24& 0.25\\
\hline
{$N_{\mathrm{He}\lesssim2.5}$}& {7675} & {20474} & {21919} & {8805} & {23767} & {19156}\\
\hline
$N_{\rm Ibn, merger}$  ($r_{\rm Ibn, merger}$ )& 622 ($9.6\times 10^{-4}$)& 1103 ($1.6\times 10^{-3}$)& 406 ($6.1\times 10^{-4}$)& 397 (5.7$\times 10^{-4}$)& 819 ( $1.2\times 10^{-3}$)& 1 ($1.4\times 10^{-6}$ ) \\
$N_{\rm He+NS/BH}$  ($r_{\rm He,CO}$)& 22965 (0.036)& 29483 (0.044)& 38719 (0.058)& 26255 (0.038)&31935 (0.046)& 29358 (0.041)\\
\hline
\end{tabular}
\caption{Number of occurrences of each event in the $10^6$ binary systems for each $(\alpha\lambda, \beta)$. $N_{\rm CCSN,bin}$ refers to the number of CCSNe in our binary models, $N_{\rm SESN}$ refers to the pre-SN He stars in our binary models expected to explode as SESNe, {$N_{\mathrm{He}\lesssim2.5}$ refers to the number of SESNe whose progenitors are He stars less massive than $2.5~M_\odot$,} $N_{\rm Ibn,merger}$ is the number of SNe Ibn caused by He stars and NS/BHs merger, and $N_{\rm He+NS/BH}$ refers to the total number of binary systems which consist of a He star and a compact object. $r_{i}$ is the corresponding fraction of event $i$ with respect to CCSNe, based on equation~(\ref{eqn:ratio}) taking into account SNe from born-single stars.}
\label{tab:result}
\end{table*}

\begin{figure*}
 \centering
 \includegraphics[width=\linewidth]{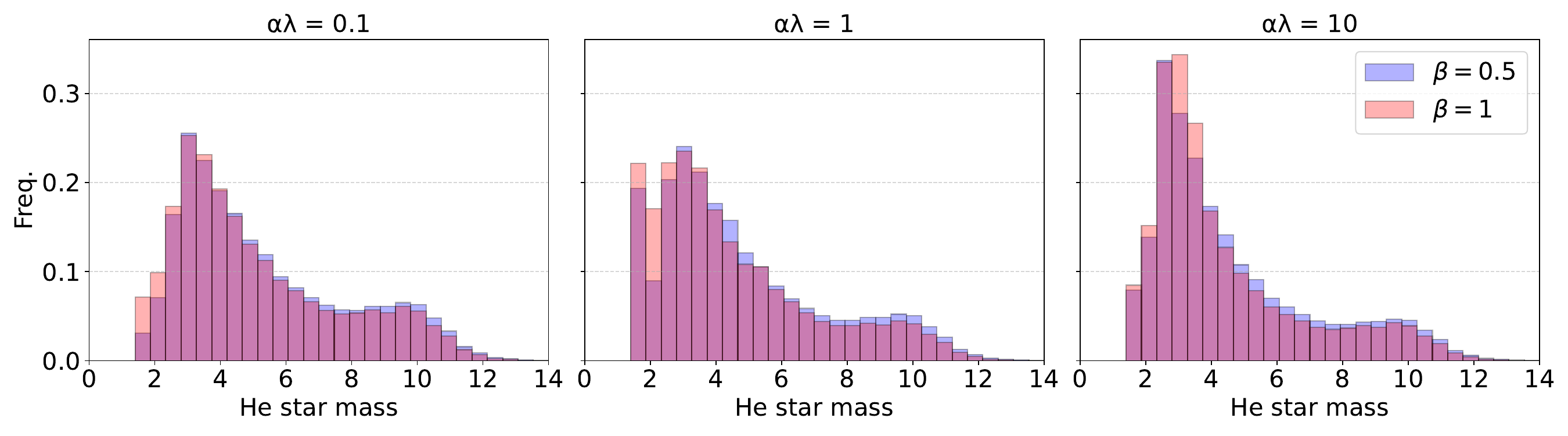}
\caption{The mass distribution of pre-explosion He stars in Figure \ref{fig:mass_sep}, for the six binary models varying $(\beta, \alpha \lambda)$. The three panels are for different $\alpha\lambda$, while the colors in each panel show results for different $\beta$.}
 \label{fig:mass_dist}
\end{figure*}

If He stars explode as CCSNe, they are expected to mostly be observed as SESNs. Table~\ref{tab:result} shows the event rates of such SESNe of binary origin $r_{\rm SESN, bin}$, derived using equation~(\ref{eqn:ratio}). In addition, massive single stars, with ZAMS masses of $>24~M_\odot$ in our model of solar metallicity, can also lose most of its hydrogen-rich envelope and explode as SESN\footnote{Note that this mass threshold is {sensitive on the choice of mass loss prescriptions as well as metallicity}, since the stellar wind mass loss rate is stronger for higher metallicity.}. The contribution from single stars to the SESNe rate can be estimated as
\begin{equation}
    r_{\rm SESN, sin} = \frac{\int_{24 M_\odot}^{100 M_\odot}\rm IMF dM}{\int_{3M_\odot}^{100 M_\odot} {\rm IMF dM} \times \frac{N_{\rm CCSN,bin}}{N_{\rm tot}}+\int_{8M_\odot}^{100 M_\odot} \rm IMF dM}\approx 9~\%.
\end{equation}
Taking this into account, the total SESN rate ($r_{\rm SESN, bin}+r_{\rm SESN,sin}$) is 20–29~\% of all CCSNe, which is in good agreement with the observationally inferred event rates of approximately $20 - 30~\%$ of CCSNe \citep[e.g.,][]{2011MNRAS.412.1522S,2011MNRAS.412.1441L,2017MNRAS.471.4381S,2020ApJ...904...35P,2020PASP..132h5002S,Ma25}.

{Recent works~\citep[e.g.,][]{2015MNRAS.451.2123T,Woosley2019} indicate that He stars with $\lesssim 2$--$2.5~M_\odot$ are unlikely to lead to CCSNe, instead leading to WDs or ECSNe. In our analysis, CCSN progenitors are identified based on a simple threshold on the CO core mass, under which He stars with $\sim 2$--$2.5~M_\odot$ are also included. Given the uncertainties on the fate of these stars, the total number of CCSNe may be overestimated. However, since the contribution $N_{\mathrm{He}\lesssim2.5}$ from such low-mass He stars is negligible compared to the overall CCSNe population ($\lesssim 5~\%$; Table \ref{tab:result}), we expect this uncertainty to only weakly affect the fraction of SESNe (and SN Ibn in the next section) with respect to CCSNe.}

The distribution of the pre-collapse He star mass is shown in Figure~\ref{fig:mass_dist}. The low-mass He stars outnumber the high-mass He stars reflecting the IMF, which has little dependence on the binary parameters $(\alpha\lambda,\beta)$. We also see that He stars with pre-SN masses of 3–4~$M_\odot$ are the most common, which will lead to SN ejecta masses of $1$--$3\ M_\odot$ assuming it leaves a NS remnant of mass $1$--$2\ M_\odot$. These align with the observed properties of SESNe, such as their typically inferred ejecta masses \citep{2011ApJ...741...97D, Lyman16, 2018A&A...609A.136T,2023ApJ...955...71R}, and the rare identification of massive single-star progenitors in pre-SN imaging \citep{2013MNRAS.436..774E, 2015PASA...32...16S}.

\subsection{Comparison with the Observed SN Ibn rate}
Here we extract the binaries that lead to SN Ibn, using our binary parameters under specific binary scenarios proposed in the literature. In this study we consider two scenarios for SN Ibn: one in which a low-mass He star undergoes CCSN in a close binary system, and another in which a He star merges with a NS or BH. Below, we describe the characteristics of each scenario and the conditions we impose on the binary parameters for each scenario. We then estimate their event rates, and compare those to the observationally inferred SN Ibn event rate of ~$1-2~\%$ of CCSNe \citep[e.g.,][]{2008MNRAS.389..113P,2022ApJ...927...25M,Ma25}.

\subsubsection{Scenario 1: Low mass He stars in Close Binaries}
Various stellar evolution modeling has suggested that the outer envelope of low mass He stars can undergo significant expansion at the late stages of evolution after core He depletion \citep[e.g.,][]{Paczynski1971,1986A&A...165...95H,1986A&A...167...61H,2013ApJ...778L..23T,2021ApJ...920L..36J,wu2022extreme,2024arXiv241209893E}. 
Especially, \cite{wu2022extreme} report that the outer envelope of He stars with masses of approximately $2.5\text{--}3.0\,M_\odot$ can undergo a non-monotonic radius evolution, with rapid expansion of up to a few $10$--$100R_\odot$ during the final months to decades of its life. Binaries consisting of such He stars can undergo mass transfer within decades before core-collapse, with the He star losing mass at an extreme rate of $\dot{M}\gtrsim 10^{-2}~M_\odot\ {\rm yr^{-1}}$. If mass transfer is not conservative, as expected for such a high mass-transfer rate, a dense, He-rich CSM forms close to the progenitor and can power a SNe Ibn upon the He star's core-collapse. 

To calculate the expected rates of SN Ibn from this channel, from our binary population we extract binaries consisting of He stars within this mass range of $2.5\text{--}3.0~M_\odot$. As can be seen from Figures~\ref{fig:mass_sep} and \ref{fig:mass_dist}, such low-mass He stars in this mass range are not rare due to the IMF favoring lower mass He stars. Additionally, for He stars of this mass range the hydrogen-rich envelope must be stripped due to binary interactions, so the existence of a binary companion is guaranteed.  

Our models do not directly predict the range of separations that can lead to the extreme CSM seen in SN Ibn, as stellar models are not evolved to such a late stage of nuclear burning in rapid BSE codes. He stars in this mass range are expected to overfill its Roche lobe for binary separations out to a few $100~R_\odot$ \citep{wu2022extreme}. However, the outer layer can exhibit complicated radial evolution via envelope stripping, making the conditions for SN Ibn less clear. Absent of a detailed grid of binary models near core-collapse, we simply calculate the expected rates of SN Ibn for various separation thresholds, assuming all low-mass He star in binaries with separations lower than a given threshold lead to SN Ibn. We consider separation thresholds out to $300\,R_\odot$, since the maximum separation for Roche-lobe overflow to occur is a typically a few times larger than the radius of the He star for most mass ratios \citep{1983ApJ...268..368E}.

\begin{figure*}
 \centering
 \includegraphics[width=\linewidth]{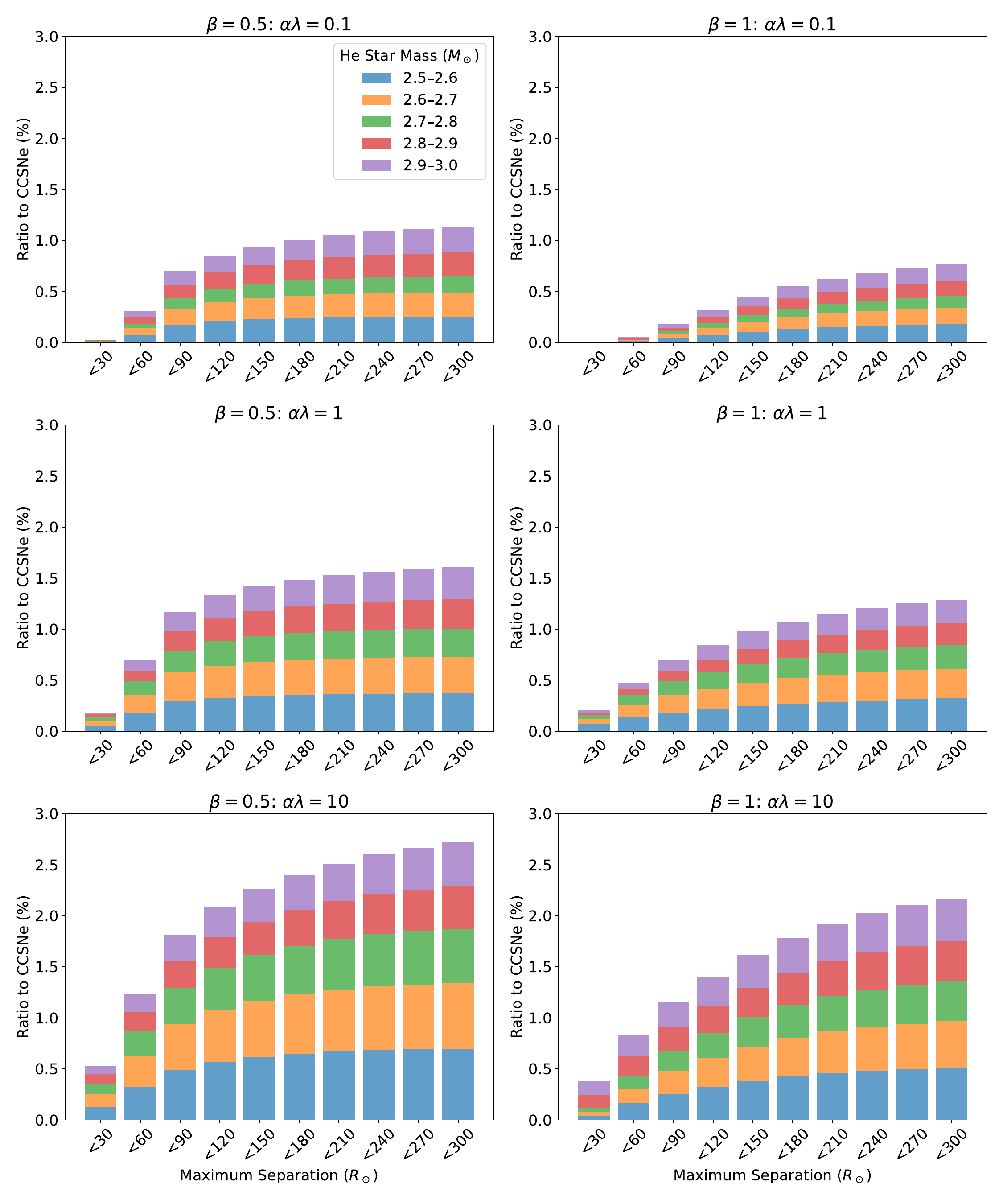}
\caption{Event rate of SN Ibn relative to CCSNe for the Low-mass (2.5--3 $M_{\odot}$) He star scenario, as a function of the maximum binary separation leading to SN Ibn. The panels show cases for the six binary models of ($\beta, \alpha \lambda)$. The colors represent different mass ranges of the He stars, separated by $0.1~M_\odot$.}
\label{fig:Ibn_rate}
\end{figure*}

Figure \ref{fig:Ibn_rate} shows the cumulative counts of binaries with He stars in the mass range of 2.5\text{--}3.0 $M_{\odot}$, as a function of maximum binary separation for SN Ibn. The horizontal axis represents the maximum separation between the binary components from 30 to 300 $R_{\odot}$, while the vertical axis displays the corresponding rate of such systems scaled by the total number of CCSNe. The color bands indicate different mass ranges for the He stars, separated by bins of $0.1~M_\odot$. All mass bins have comparable numbers of He star binaries, for all separations and binary parameters $(\beta,\alpha\lambda)$.

The fraction of binary systems including low mass He star relative to the total CCSN population is around 1\text{--}3\% across the values of ($\beta, \alpha \lambda$), suggesting that He stars with mass of 2.5\text{--}3.0 $M_\odot$ alone can account for the entire observed rate of SNe Ibn. This result supports the low-mass He star scenario for producing SN Ibn. For $\beta = 0.5$, the rate of SNe Ibn relative to the total CCSNe is slightly higher than in the case of $\beta = 1$. This is likely because the binary separation tends to be smaller for $\beta = 0.5$, making it easier to strip the envelope and thus increasing the number of He stars compared to the $\beta = 1$ case. The fraction also appears to be higher for larger values of $\alpha \lambda$, which may be because, within the separation range of $\lesssim$ a few 100 $R_\odot$ considered here, smaller $\alpha \lambda$ values are more likely to lead to binary mergers, reducing the number of surviving binary systems.

It is important to note that this scenario is realized in two kinds of binary populations: one where the more massive star on the ZAMS is the SN Ibn progenitor, and another where the initially less massive one becomes the SN Ibn progenitor. In the former case, the companion stars are typically main sequence (MS) stars or other hydrogen-rich stars, which survive after the explosion of the He star. In the latter case, the companions of these exploding He stars are generally compact objects, interestingly including (and potentially dominated by) white dwarfs. The characteristics of these companion stars will be discussed in detail in Section \ref{sec:discussion}.

{We note that the chosen threshold of the He star mass range is approximate. This is in part motivated by stellar evolution theory, where He stars with masses less than $2.5~M_\odot$ are known to be difficult to explode~\citep[e.g.,][]{2015MNRAS.451.2123T,Woosley2019} while those with masses well exceeding $3~M_\odot$ are not expected to rapidly expand in the end stages of stellar evolution. Nevertheless, we find that our estimated rates of 1--3\% are not strongly affected by the chosen mass thresholds, as demonstrated in Appendix \ref{sec:app_massrange}.}

\subsubsection{Scenario 2: Mergers of He star + NS/BH binaries}
Some studies suggest that the conditions of SNe Ibn could be met by binary systems composed of a He star and NS/BH undergoing merger \citep[e.g.,][]{2012ApJ...752L...2C,2022ApJ...932...84M,2023ApJ...955..125T}. The binary interaction preceding merger forms the dense CSM, and an explosion is triggered when the NS/BH merges with the He star. {In this case, the explosion in this scenario is not caused by a canonical CCSN, but the observational signatures are expected to resemble interaction-powered SNe.} While the properties of the CSM and the final explosion are theoretically uncertain \citep[e.g.,][]{Fryer98,Zhang01,Schroder20}, such mergers have previously been proposed for a Type Ibn SN 2023fyq, which showed a years-long precursor emission that gradually brightens to the terminal explosion \citep{2024ApJ...977..254D,2024OJAp....7E..82T}.

In this scenario, a NS or BH is first formed via a CCSN, and in the more likely case of a NS formation the binary system must survive the associated natal kick. This requirement significantly limits the number of systems that can realize both envelope stripping of the secondary (to form a He star) and a merger afterwards. This is especially the case when compared to the low-mass He star scenario, that does not require a prior SN event.

To investigate the event rate and progenitor properties of such systems, we extracted binary systems consisting of a He star and a NS or BH that undergo merger. The method to extract such binaries is summarized in Appendix~\ref{sec:app_merger}, and the total number of these systems is indicated as $N_{\rm Ibn,merger}$ in Table~\ref{tab:result}. The event rate relative to CCSNe, $r_{\rm Ibn, merger}$, can be similarly derived from equation~(\ref{eqn:ratio}), and is also listed in Table~\ref{tab:result}. We find that the He star + NS/BH merger scenario struggles to explain the observed SN Ibn rate, with predicted event rates of $\lesssim 0.1\%$ of CCSNe that is an order of magnitude lower than the observed rate. 

However, this does not necessarily mean that the merger scenario is entirely ruled out as a progenitor pathway for SNe Ibn. This is because the rapid binary population synthesis code used in this study does not include certain physical processes that could potentially increase the number of mergers. For example, in the post-CE scenario of \cite{2022ApJ...932...84M} the key process driving the merger is the angular momentum exchange between the binary and the post-CE circumbinary disk, that leads to a further decrease in the binary separation \citep[see also][]{2023ApJ...955..125T,2024A&A...688A..87W}. The rapid binary population synthesis codes do not take into account the effects of the (bound) material following a successful CE ejection. Furthermore, our binary models based on \cite{2002MNRAS.329..897H} potentially underestimate the radius expansion of stripped stars at the lowest masses in the late stages of its evolution \citep[e.g.,][Figure 6]{2020A&A...637A...6L}, which can lead to lower rates of mergers.
 
In fact, we find that there are approximately 20,000-30,000 He star + NS/BH binary systems that do not merge in our BSE calculation, corresponding to about a few~\% of all CCSNe if they all lead to SN Ibn (see Table~\ref{tab:result}). If additional physical processes like above are taken into account, it is possible that the number of SNe Ibn originating from mergers could significantly increase. We leave such considerations for future work.
\section{Discussion}
\label{sec:discussion}
\subsection{Features of the Companion Stars}
\label{sec:companions}
\begin{figure*}
 \centering
 \includegraphics[width=\linewidth]{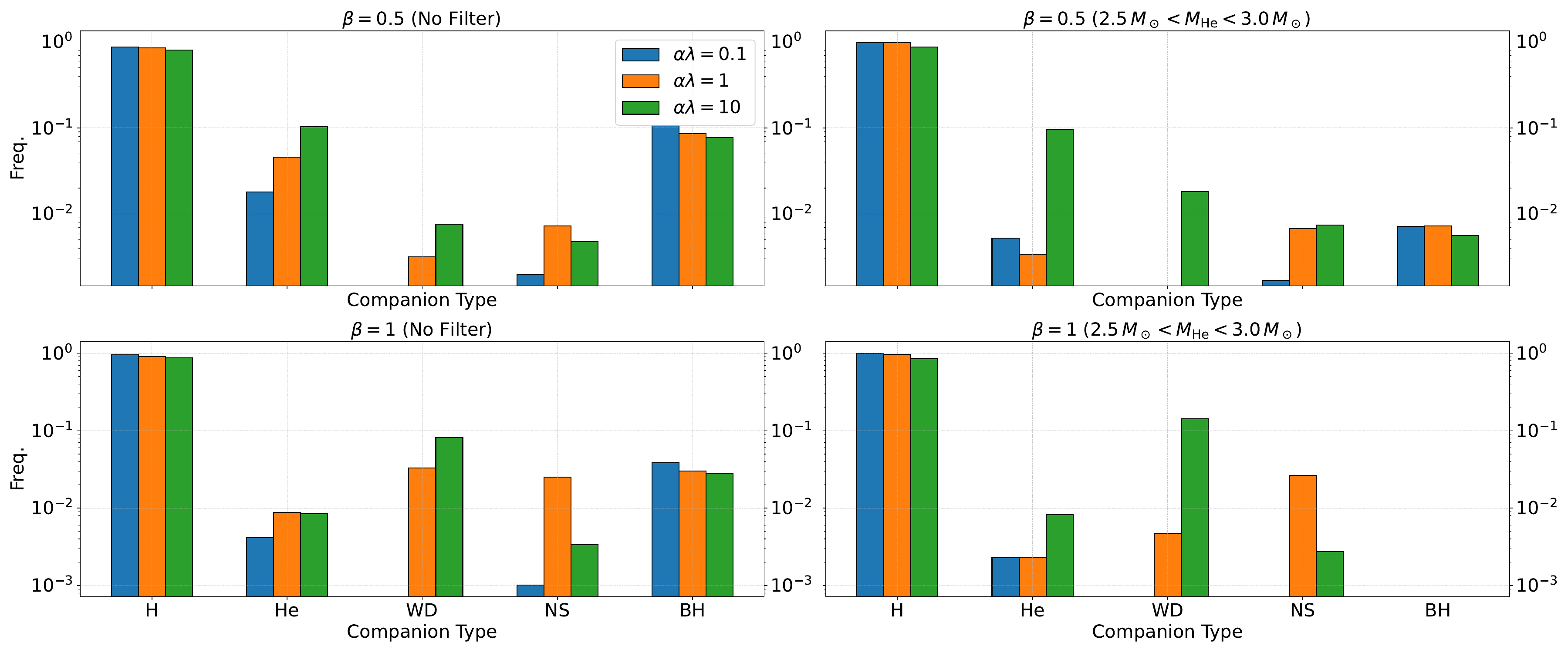}
\caption{The normalized distribution of companion star types for exploding He stars for different binary parameters ($\beta,\alpha\lambda$). The left panel shows the types of companions for all He stars with no mass restrictions, while the right panel displays the companion types for He stars with masses in the range $2.5$--$3.0\,M_\odot$. "H" refers to the stars with hydrogen-rich envelopes such as MS stars and giants, which correspond to the stellar types 0 to 6 in the classification of \citet{2002MNRAS.329..897H}. "He" refers to the He stars (7, 8, 9 in \citealt{2002MNRAS.329..897H}). "WD" includes the HeWD, COWD and ONeWD (10, 11, 12 in \citealt{2002MNRAS.329..897H}).}
 \label{fig:companion_type}
\end{figure*}

In contrast to the merger scenario, the low mass He star scenario predicts that the SN leaves behind a surviving companion star. Thus the most straightforward observational test to distinguish these formation scenarios is to search for these surviving companions for nearby SN Ibn events. Here we discuss the properties of the companion stars, mainly focusing on the low-mass He star channel.

The types of companion stars predicted from our binary models are summarized in Figure~\ref{fig:companion_type}, with the left panels for all exploding He stars and the right panels for He stars in the mass range $2.5$--$3~M_\odot$ relevant for SN Ibn. For all cases, almost all ($\gtrsim 90~\%$) of the leftover companion stars are stars with hydrogen-rich envelopes, predominantly being main sequence stars. Interestingly, WD companions are also non-negligible ($\sim 10~\%$) for some binary parameters, especially for larger values of $\alpha \lambda$. {These features are also seen in previous binary population synthesis modeling for the general population of SESNe~\citep[][]{2017ApJ...842..125Z}, where the companions are mainly main sequence stars of $\sim 10~M_\odot$, with compact objects occupying a small fraction (1--10\% for different model assumptions).}

\begin{figure*}
 \centering
 \includegraphics[width=\linewidth]{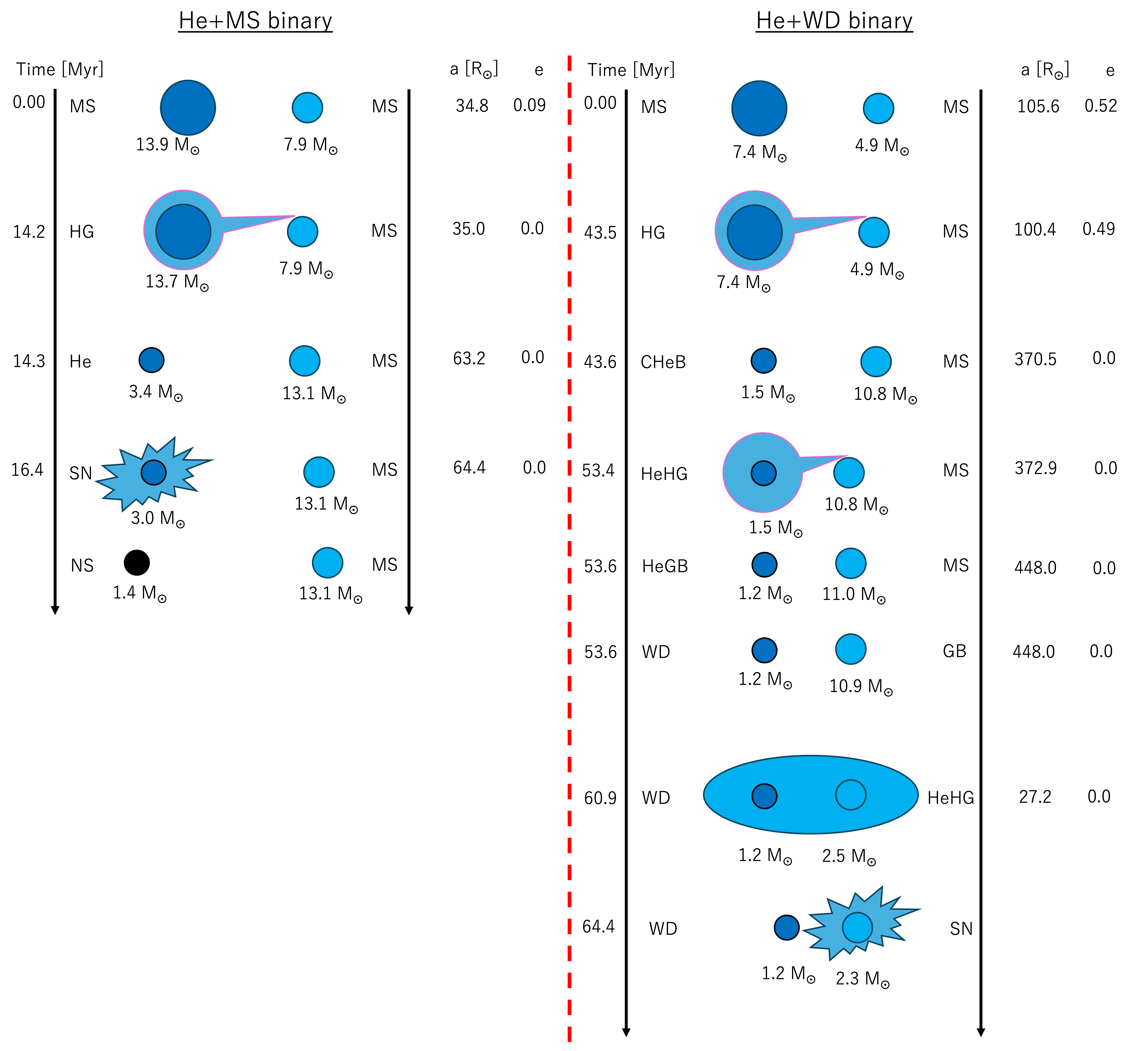}
\caption{Representative evolutionary paths of systems in which a low mass He star with $2.5\text{--}3~M_\odot$ is accompanied by a MS star (left) or a WD (right). The left case appears for all cases of ($\beta,\alpha\lambda$), and the specific numbers are for a binary model  with $\alpha \lambda = 0.1$, $\beta = 1$. The right case has a longer lifetime and appears more frequently for higher $\alpha\lambda$, and the specific numbers are for a binary in the model with $\alpha \lambda = 10$, $\beta = 1$. }
 \label{fig:evo_path}
\end{figure*}

The representative evolution paths for these two scenarios are schematically shown in Figure~\ref{fig:evo_path}. The left panel shows the case in which a MS star remains after the SN, whereas the right panel represents the case in which a WD is left behind. While the left scenario generally happens regardless of the values of $\beta$ and $\alpha\lambda$ (albeit with slightly different rates), the right scenario is more rare, and exclusively occurs for larger $\alpha\lambda$ and $\beta$. The rarity is because of the tuning of the primary to be born as both low enough mass to  become a WD, while at the same time high enough mass such that the secondary accretes the right amount of mass to lead to a $2.5$--$3\ M_\odot$ He star (the requirement of mass reversal favors large $\beta$). Furthermore, a larger $\alpha\lambda$ is required for the final CE event that forms the He star to be successful, due to the large mass ratio between a WD and a $\sim 10~M_\odot$ donor star. Such a large $\alpha\lambda$ is nevertheless not impossible, and the (effective) value of $\alpha \lambda$ may become as large as this when the donor star expands and triggers a CE during its Hertzsprung gap phase \citep[e.g.,][]{2019ApJ...883L..45F,2022ApJ...937L..42H}.

\begin{figure*}
 \centering
 \includegraphics[width=\linewidth]{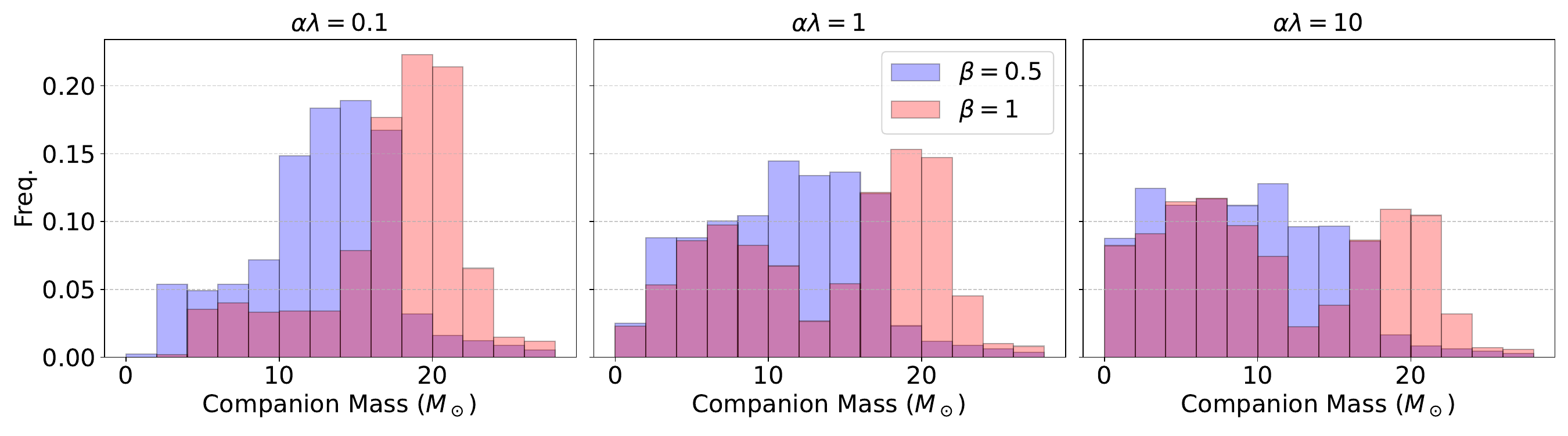}
\caption{The normalized mass distribution of a MS companion star of 2.5\text{--}3 $M_\odot$ He stars, at the moment of the explosion of the He star.  The three panels are for different $\alpha\lambda$, while the colors in each panel show results for different $\beta$.}
 \label{fig:companion_mass}
\end{figure*}

The prevalence of surviving stellar companions agrees well with previous post-SN observations of nearby SN Ibn, such as SN 2006jc~\citep[e.g.,][]{2016ApJ...833..128M,Sun20}. The expected distributions of the MS companion's masses are summarized in Figure~\ref{fig:companion_mass}. The masses of surviving MS companions are distributed nearly uniformly up to $\sim$ 15--20 $M_\odot$ and heavier for higher $\beta$, as generally the companion gains mass due to stable mass transfer from the primary that becomes the He star. 

\begin{figure*}
 \centering
 \includegraphics[width=\linewidth]{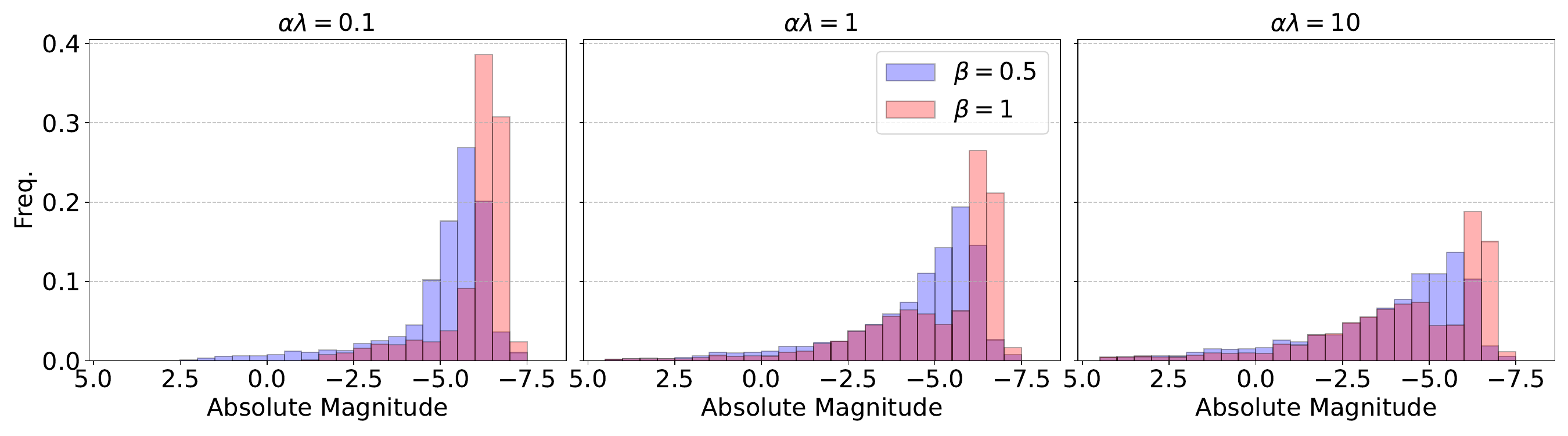}
\caption{Same as Figure \ref{fig:companion_mass}, but a distribution of the absolute magnitude in near-UV ($\approx 300$ nm; see Section \ref{sec:companions} for details).}
 \label{fig:abs_mag}
\end{figure*}

Our binary models may serve as a benchmark for future work to estimate the expected detectability of the companions in nearby SN Ibn. To demonstrate this, we show in Figure~\ref{fig:abs_mag} the distribution of the companion's absolute magnitude in near-UV (F300X filter in {\it HST}). We have used the synthetic photometry provided by the MIST (MESA Isochrones and Stellar Tracks) models~\citep[][]{MIST1,MIST2}, and considered stars in the middle of the main sequence where $X_{\rm H}=Y_{\rm He}$ at the center. We find that the majority of the MS companions are brighter than $\sim -5$ mag, with a brighter peak for larger $\beta$. For a detection limit of 26~mag with \textit{HST}'s Wide Field Camera 3 (as was the case for the companion of SN 2006jc;~\citealt{Sun20}), we expect these to be observable for sufficiently nearby SNe Ibn occurring within $\sim 15$--$20$ Mpc. We note that the observational appearance of the companion can potentially be altered due to interaction with the SN ejecta \citep{Wheeler75,Hirai18,2021MNRAS.505.2485O}, generally making the star brighter and redder for a thermal timescale. While the magnitude and timescale of the brightening depend also on the binary separation and the SN energy, we expect this to generally improve the prospects for finding these companions in the optical and near-UV. For SN 2006jc, \cite{Sun20} constrains the companion to be either a $\sim 12~M_\odot$ star that just evolved off the MS, or a lower-mass MS star that became brightened by interaction with the SN ejecta.

While surviving WD companions would be difficult to observe, the possible existence of such progenitor channel may be consistent with observations of some SNe Ibn identified in regions devoid of star-formation, such as PS1-12sk \citep[e.g.,][]{2013ApJ...769...39S,2019ApJ...871L...9H}. The offset from the candidate star-forming region implies that the progenitors have experienced significant delay times ($\gtrsim 100$~Myr) after star formation. Such observations are difficult to explain with typical massive-star progenitors, whose lifetimes are $\lesssim 10$~Myr. Among the possible progenitor channels, binaries composed of a WD and a low-mass He star identified in this study may offer a potential explanation for such events, as they can in principle survive for $\sim 100$~Myr before the SN Ibn occurs. This much longer delay time is mainly due to the lower mass of the primary, with the secondary accreting just the right mass to form a He star of $2.5$--$3~M_\odot$ after the CE with the WD remnant. {Such delay times of $\sim 100$ Myrs has also been suggested for canonical CCSNe as well, to predominantly occur in binary systems with similar primary masses as ours \citep[e.g.,][]{2003NewA....8..415D,2017A&A...601A..29Z}}

In this scenario of a WD companion, after the CE ejection the He star may further provide He-rich material onto the WD via stable mass transfer, potentially triggering nova-like outbursts prior to the SN Ibn explosion. Recently \citet{2025arXiv250312586H} report that if the mass transfer rate is $\lesssim 10^{-5}M_\odot\mathrm{yr}^{-1}$, He can be steadily accreted onto the WD, potentially leading to recurrent He-novae. Given that these He stars form $\sim 10^6$ years before the final SN, these events may occur as early as $\sim 10^5$--$10^6$ years before SN Ibn. Since these comprise $<10~\%$ of SN Ibn (largest for $\alpha\lambda=10$), and adopting a Galactic CCSN rate of a few per century, we estimate that there may be at most dozens of such systems in our Galaxy.


\subsection{Avenues for Future Work}
In this study, we do not follow the detailed binary evolution up to the onset of CCSN. Instead, we assume that all low-mass He stars within a specific mass range in binary systems could form dense He-rich CSM and eventually lead to SNe Ibn.
However, in order to estimate the event rate more accurately, it is necessary to incorporate the physical processes responsible for He-rich CSM formation and to apply more realistic criteria for the relevant mass range. Recent studies using stellar evolution codes such as MESA \citep[e.g.,][]{wu2022extreme,2024arXiv241209893E,2025arXiv250301993G} have investigated the evolution of low-mass He stars in binary systems in detail, and at much later stages than captured in this study using rapid binary population synthesis. These works place constraints on which binary configurations and He-star mass ranges can lead to the formation of dense CSM and potentially result in SNe Ibn.
Further refining these constraints and incorporating them into binary population synthesis models will allow us to more accurately characterize the progenitor systems and overall properties of SNe Ibn.

Additionally, metallicity may influence the rates of SNe Ibn as well as its observational appearance. Since lower metallicity leads to more compact stellar structures, Roche-lobe overflow becomes less frequent. As a result, the number of binary systems in which an exploding He star undergoes mass transfer may decrease, potentially reducing the overall number of SNe Ibn. On the other hand, lower metallicity tends to make the star expand less during the Hertzsprung gap and helps avoid stellar mergers during the CE phase following case B mass transfer \citep{Belczynski2010ApJ...715L.138B}, hence increasing the number of surviving binaries containing He stars \citep[see also][]{Kinugawa2017ApJ...849L..29K}. Finally, when hydrogen is stripped via mass transfer the donor tends to leave more residual hydrogen for lower metallicities \citep{2017A&A...608A..11G,2020A&A...637A...6L}. This can affect the hydrogen content of SESNe and potentially the SN Ibn rate. Overall, the net effect remains unclear and is left for future work. As stellar winds are weaker at lower metallicity \citep[e.g.][]{Vink2001A&A...369..574V}, binary interactions play an even more critical role in producing stripped stars under such conditions. Therefore, it is important to perform more detailed estimates in future work to understand the metallicity dependence of SNe Ibn rates and progenitor types. 

We also assume 1.44~$M_{\odot}$ for the lower limit of the CO core mass for core-collapse, and 250~$\rm km~s^{-1}$ for the natal kick velocity dispersion for NSs. {While we neglected the contribution of ECSNe case,} the natal kicks of ECSNe are suggested to be smaller than the canonical value assumed in this work ($\sim 30\rm ~km~s^{-1}$) \citep{2002ApJ...571L..37P,2004ApJ...612.1044P,2019ApJ...882...93H}. If we consider the effect of ECSN progenitors, the He star–NS binary systems that lead to SN Ibn (as considered in e.g., \citealt{wu2022extreme,2024OJAp....7E..82T,2024ApJ...977..254D}) might increase. While the observationally constrained rate of ECSNe is not large \citep[$0.6$--$8.5\%$ of CCSNe;][]{2021NatAs...5..903H}, the ZAMS masses relevant for ECSNe are close to those required to produce low-mass He stars relevant for SN Ibn.

Finally, while we have focused on understanding the binary systems leading to SN Ibn, we have not made direct predictions for the SN itself. When more detailed theoretical binary modeling of low-mass He stars become available, their stellar structures can provide insights into the ejecta mass and explosion energy realized upon core-collapse of the He star. Light curve modeling of these explosions and comparisons with observations of SNe Ibn can also be an important task \citep[e.g.,][]{2024arXiv241209893E,2025arXiv250308790H}.

\section{Conclusion}
\label{sec:conclusion}
In this paper, we performed rapid binary population synthesis calculations to estimate the event rate of SNe Ibn, considering two binary scenarios proposed for these SNe: extreme mass transfer from low-mass He stars and merger between He stars and NS/BHs. Our binary models containing He stars (Figure \ref{fig:mass_sep}) cover a broad region of the mass-separation parameter space where He stars are expected to be produced by various stellar/binary evolution processes, and correctly predicts the observed event rate of SESN.

From our binary models, we find that while the low-mass He star scenario can sufficiently explain the rate of SN Ibn, the merger scenario struggles to account for the observed rate by at least an order of magnitude. We note however, that this does not necessarily imply that the binary merger scenarios are completely discarded, due to key physical processes leading to merger potentially lacking in the binary population synthesis modeling.

Furthermore, under the low-mass He star scenario we made predictions for the companion star left after the SN Ibn explosion. We find that a MS star is left behind in most cases ($\gtrsim 90\%$), with a broad range of masses up to $\sim 20$ $M_\odot$. In rare cases ($\lesssim 10\%$) a WD may be left behind, which has not yet been directly observed but can potentially accomodate for the existence of SNe Ibn found in regions with no obvious star formation.

Due to the rarity of SN Ibn, the samples of nearby SN Ibn with detailed constraints on the binary properties are currently scarce. Our binary models can be an important step for detailed future predictions of these post-explosion observations, which can provide further observational tests to these scenarios.

\section*{Acknowledgements}
We thank Toshikazu Shigeyama, James Fuller and Ilya Mandel for valuable discussions. We also thank the referee for insightful comments and suggestions that improve the clarity of this paper. Takatoshi Ko is supported by JSPS KAKENHI grant No. 24KJ0672. This work was made possible through the support of the Enrico Fermi Fellowships led by the Center for Spacetime and the Quantum, and supported by Grant ID \#63132 from the John Templeton Foundation. The opinions expressed in this publication are those of the author(s) and do not necessarily reflect the views of the John Templeton Foundation or those of the Center for Spacetime and the Quantum.
Tomoya Kinugawa is supported by JSPS KAKENHI grant No. 23H04893.
Daichi Tsuna is supported by the Sherman Fairchild Postdoctoral Fellowship at the California Institute of Technology.
Yuki Takei acknowledges support from the JSPS KAKENHI grant No.~23H04900.

 \section*{Data Availability}
All data underlying this article are available on reasonable request to the corresponding author.



\bibliographystyle{mnras}
\bibliography{reference} 

\begin{thebibliography}{}
\makeatletter
\relax
\def\mn@urlcharsother{\let\do\@makeother \do\$\do\&\do\#\do\^\do\_\do\%\do\~}
\def\mn@doi{\begingroup\mn@urlcharsother \@ifnextchar [ {\mn@doi@} {\mn@doi@[]}}
\def\mn@doi@[#1]#2{\def\@tempa{#1}\ifx\@tempa\@empty \href {http://dx.doi.org/#2} {doi:#2}\else \href {http://dx.doi.org/#2} {#1}\fi \endgroup}
\def\mn@eprint#1#2{\mn@eprint@#1:#2::\@nil}
\def\mn@eprint@arXiv#1{\href {http://arxiv.org/abs/#1} {{\tt arXiv:#1}}}
\def\mn@eprint@dblp#1{\href {http://dblp.uni-trier.de/rec/bibtex/#1.xml} {dblp:#1}}
\def\mn@eprint@#1:#2:#3:#4\@nil{\def\@tempa {#1}\def\@tempb {#2}\def\@tempc {#3}\ifx \@tempc \@empty \let \@tempc \@tempb \let \@tempb \@tempa \fi \ifx \@tempb \@empty \def\@tempb {arXiv}\fi \@ifundefined {mn@eprint@\@tempb}{\@tempb:\@tempc}{\expandafter \expandafter \csname mn@eprint@\@tempb\endcsname \expandafter{\@tempc}}}

\bibitem[\protect\citeauthoryear{{Abt}}{{Abt}}{1983}]{1983ARA&A..21..343A}
{Abt} H.~A.,  1983, \mn@doi [\araa] {10.1146/annurev.aa.21.090183.002015}, \href {https://ui.adsabs.harvard.edu/abs/1983ARA&A..21..343A} {21, 343}

\bibitem[\protect\citeauthoryear{{Antoni} \& {Quataert}}{{Antoni} \& {Quataert}}{2023}]{Antoni23}
{Antoni} A.,  {Quataert} E.,  2023, \mn@doi [\mnras] {10.1093/mnras/stad2328}, \href {https://ui.adsabs.harvard.edu/abs/2023MNRAS.525.1229A} {525, 1229}

\bibitem[\protect\citeauthoryear{{Belczynski}, {Taam}, {Kalogera}, {Rasio}  \& {Bulik}}{{Belczynski} et~al.}{2007}]{2007ApJ...662..504B}
{Belczynski} K.,  {Taam} R.~E.,  {Kalogera} V.,  {Rasio} F.~A.,   {Bulik} T.,  2007, \mn@doi [\apj] {10.1086/513562}, \href {https://ui.adsabs.harvard.edu/abs/2007ApJ...662..504B} {662, 504}

\bibitem[\protect\citeauthoryear{{Belczynski}, {Dominik}, {Bulik}, {O'Shaughnessy}, {Fryer}  \& {Holz}}{{Belczynski} et~al.}{2010}]{Belczynski2010ApJ...715L.138B}
{Belczynski} K.,  {Dominik} M.,  {Bulik} T.,  {O'Shaughnessy} R.,  {Fryer} C.,   {Holz} D.~E.,  2010, \mn@doi [\apjl] {10.1088/2041-8205/715/2/L138}, \href {https://ui.adsabs.harvard.edu/abs/2010ApJ...715L.138B} {715, L138}

\bibitem[\protect\citeauthoryear{{Brandt} \& {Podsiadlowski}}{{Brandt} \& {Podsiadlowski}}{1995}]{1995MNRAS.274..461B}
{Brandt} N.,  {Podsiadlowski} P.,  1995, \mn@doi [\mnras] {10.1093/mnras/274.2.461}, \href {https://ui.adsabs.harvard.edu/abs/1995MNRAS.274..461B} {274, 461}

\bibitem[\protect\citeauthoryear{{Burrows}, {Wang}  \& {Vartanyan}}{{Burrows} et~al.}{2024}]{Burrows24}
{Burrows} A.,  {Wang} T.,   {Vartanyan} D.,  2024, \mn@doi [arXiv e-prints] {10.48550/arXiv.2412.07831}, \href {https://ui.adsabs.harvard.edu/abs/2024arXiv241207831B} {p. arXiv:2412.07831}

\bibitem[\protect\citeauthoryear{{Chevalier}}{{Chevalier}}{2012}]{2012ApJ...752L...2C}
{Chevalier} R.~A.,  2012, \mn@doi [\apjl] {10.1088/2041-8205/752/1/L2}, \href {https://ui.adsabs.harvard.edu/abs/2012ApJ...752L...2C} {752, L2}

\bibitem[\protect\citeauthoryear{{Choi}, {Dotter}, {Conroy}, {Cantiello}, {Paxton}  \& {Johnson}}{{Choi} et~al.}{2016}]{MIST2}
{Choi} J.,  {Dotter} A.,  {Conroy} C.,  {Cantiello} M.,  {Paxton} B.,   {Johnson} B.~D.,  2016, \mn@doi [\apj] {10.3847/0004-637X/823/2/102}, \href {https://ui.adsabs.harvard.edu/abs/2016ApJ...823..102C} {823, 102}

\bibitem[\protect\citeauthoryear{{Davis} et~al.,}{{Davis} et~al.}{2023}]{2023MNRAS.523.2530D}
{Davis} K.~W.,  et~al., 2023, \mn@doi [\mnras] {10.1093/mnras/stad1433}, \href {https://ui.adsabs.harvard.edu/abs/2023MNRAS.523.2530D} {523, 2530}

\bibitem[\protect\citeauthoryear{{De Donder} \& {Vanbeveren}}{{De Donder} \& {Vanbeveren}}{2003}]{2003NewA....8..415D}
{De Donder} E.,  {Vanbeveren} D.,  2003, \mn@doi [\na] {10.1016/S1384-1076(03)00002-2}, \href {https://ui.adsabs.harvard.edu/abs/2003NewA....8..415D} {8, 415}

\bibitem[\protect\citeauthoryear{{Dessart}, {Hillier}  \& {Kuncarayakti}}{{Dessart} et~al.}{2022}]{2022A&A...658A.130D}
{Dessart} L.,  {Hillier} D.~J.,   {Kuncarayakti} H.,  2022, \mn@doi [\aap] {10.1051/0004-6361/202142436}, \href {https://ui.adsabs.harvard.edu/abs/2022A&A...658A.130D} {658, A130}

\bibitem[\protect\citeauthoryear{{Dong} et~al.,}{{Dong} et~al.}{2024}]{2024ApJ...977..254D}
{Dong} Y.,  et~al., 2024, \mn@doi [\apj] {10.3847/1538-4357/ad8de6}, \href {https://ui.adsabs.harvard.edu/abs/2024ApJ...977..254D} {977, 254}

\bibitem[\protect\citeauthoryear{{Dotter}}{{Dotter}}{2016}]{MIST1}
{Dotter} A.,  2016, \mn@doi [\apjs] {10.3847/0067-0049/222/1/8}, \href {https://ui.adsabs.harvard.edu/abs/2016ApJS..222....8D} {222, 8}

\bibitem[\protect\citeauthoryear{{Drout} et~al.,}{{Drout} et~al.}{2011}]{2011ApJ...741...97D}
{Drout} M.~R.,  et~al., 2011, \mn@doi [\apj] {10.1088/0004-637X/741/2/97}, \href {https://ui.adsabs.harvard.edu/abs/2011ApJ...741...97D} {741, 97}

\bibitem[\protect\citeauthoryear{{Duquennoy} \& {Mayor}}{{Duquennoy} \& {Mayor}}{1991}]{1991A&A...248..485D}
{Duquennoy} A.,  {Mayor} M.,  1991, \aap, \href {https://ui.adsabs.harvard.edu/abs/1991A&A...248..485D} {248, 485}

\bibitem[\protect\citeauthoryear{{Eggleton}}{{Eggleton}}{1983}]{1983ApJ...268..368E}
{Eggleton} P.~P.,  1983, \mn@doi [\apj] {10.1086/160960}, \href {https://ui.adsabs.harvard.edu/abs/1983ApJ...268..368E} {268, 368}

\bibitem[\protect\citeauthoryear{{Eldridge}, {Izzard}  \& {Tout}}{{Eldridge} et~al.}{2008}]{2008MNRAS.384.1109E}
{Eldridge} J.~J.,  {Izzard} R.~G.,   {Tout} C.~A.,  2008, \mn@doi [\mnras] {10.1111/j.1365-2966.2007.12738.x}, \href {https://ui.adsabs.harvard.edu/abs/2008MNRAS.384.1109E} {384, 1109}

\bibitem[\protect\citeauthoryear{{Eldridge}, {Fraser}, {Smartt}, {Maund}  \& {Crockett}}{{Eldridge} et~al.}{2013}]{2013MNRAS.436..774E}
{Eldridge} J.~J.,  {Fraser} M.,  {Smartt} S.~J.,  {Maund} J.~R.,   {Crockett} R.~M.,  2013, \mn@doi [\mnras] {10.1093/mnras/stt1612}, \href {https://ui.adsabs.harvard.edu/abs/2013MNRAS.436..774E} {436, 774}

\bibitem[\protect\citeauthoryear{{Ercolino}, {Jin}, {Langer}  \& {Dessart}}{{Ercolino} et~al.}{2024}]{2024arXiv241209893E}
{Ercolino} A.,  {Jin} H.,  {Langer} N.,   {Dessart} L.,  2024, \mn@doi [arXiv e-prints] {10.48550/arXiv.2412.09893}, \href {https://ui.adsabs.harvard.edu/abs/2024arXiv241209893E} {p. arXiv:2412.09893}

\bibitem[\protect\citeauthoryear{{Filippenko}}{{Filippenko}}{1997}]{1997ARA&A..35..309F}
{Filippenko} A.~V.,  1997, \mn@doi [\araa] {10.1146/annurev.astro.35.1.309}, \href {https://ui.adsabs.harvard.edu/abs/1997ARA&A..35..309F} {35, 309}

\bibitem[\protect\citeauthoryear{{Foley}, {Smith}, {Ganeshalingam}, {Li}, {Chornock}  \& {Filippenko}}{{Foley} et~al.}{2007}]{2007ApJ...657L.105F}
{Foley} R.~J.,  {Smith} N.,  {Ganeshalingam} M.,  {Li} W.,  {Chornock} R.,   {Filippenko} A.~V.,  2007, \mn@doi [\apjl] {10.1086/513145}, \href {https://ui.adsabs.harvard.edu/abs/2007ApJ...657L.105F} {657, L105}

\bibitem[\protect\citeauthoryear{{Fragos}, {Andrews}, {Ramirez-Ruiz}, {Meynet}, {Kalogera}, {Taam}  \& {Zezas}}{{Fragos} et~al.}{2019}]{2019ApJ...883L..45F}
{Fragos} T.,  {Andrews} J.~J.,  {Ramirez-Ruiz} E.,  {Meynet} G.,  {Kalogera} V.,  {Taam} R.~E.,   {Zezas} A.,  2019, \mn@doi [\apjl] {10.3847/2041-8213/ab40d1}, \href {https://ui.adsabs.harvard.edu/abs/2019ApJ...883L..45F} {883, L45}

\bibitem[\protect\citeauthoryear{{Fryer} \& {Woosley}}{{Fryer} \& {Woosley}}{1998}]{Fryer98}
{Fryer} C.~L.,  {Woosley} S.~E.,  1998, \mn@doi [\apjl] {10.1086/311493}, \href {https://ui.adsabs.harvard.edu/abs/1998ApJ...502L...9F} {502, L9}

\bibitem[\protect\citeauthoryear{{Gal-Yam} et~al.,}{{Gal-Yam} et~al.}{2022}]{2022Natur.601..201G}
{Gal-Yam} A.,  et~al., 2022, \mn@doi [\nat] {10.1038/s41586-021-04155-1}, \href {https://ui.adsabs.harvard.edu/abs/2022Natur.601..201G} {601, 201}

\bibitem[\protect\citeauthoryear{{Gilkis}, {Laplace}, {Arcavi}, {Shenar}  \& {Schneider}}{{Gilkis} et~al.}{2025}]{2025arXiv250301993G}
{Gilkis} A.,  {Laplace} E.,  {Arcavi} I.,  {Shenar} T.,   {Schneider} F.,  2025, \mn@doi [arXiv e-prints] {10.48550/arXiv.2503.01993}, \href {https://ui.adsabs.harvard.edu/abs/2025arXiv250301993G} {p. arXiv:2503.01993}

\bibitem[\protect\citeauthoryear{{G{\"o}tberg}, {de Mink}  \& {Groh}}{{G{\"o}tberg} et~al.}{2017}]{2017A&A...608A..11G}
{G{\"o}tberg} Y.,  {de Mink} S.~E.,   {Groh} J.~H.,  2017, \mn@doi [\aap] {10.1051/0004-6361/201730472}, \href {https://ui.adsabs.harvard.edu/abs/2017A&A...608A..11G} {608, A11}

\bibitem[\protect\citeauthoryear{{Habets}}{{Habets}}{1986a}]{1986A&A...165...95H}
{Habets} G.~M.~H.~J.,  1986a, \aap, \href {https://ui.adsabs.harvard.edu/abs/1986A&A...165...95H} {165, 95}

\bibitem[\protect\citeauthoryear{{Habets}}{{Habets}}{1986b}]{1986A&A...167...61H}
{Habets} G.~M.~H.~J.,  1986b, \aap, \href {https://ui.adsabs.harvard.edu/abs/1986A&A...167...61H} {167, 61}

\bibitem[\protect\citeauthoryear{{Haynie}, {Wu}, {Piro}  \& {Fuller}}{{Haynie} et~al.}{2025}]{2025arXiv250308790H}
{Haynie} A.,  {Wu} S.~C.,  {Piro} A.~L.,   {Fuller} J.,  2025, \mn@doi [arXiv e-prints] {10.48550/arXiv.2503.08790}, \href {https://ui.adsabs.harvard.edu/abs/2025arXiv250308790H} {p. arXiv:2503.08790}

\bibitem[\protect\citeauthoryear{{Heggie}}{{Heggie}}{1975}]{1975MNRAS.173..729H}
{Heggie} D.~C.,  1975, \mn@doi [\mnras] {10.1093/mnras/173.3.729}, \href {https://ui.adsabs.harvard.edu/abs/1975MNRAS.173..729H} {173, 729}

\bibitem[\protect\citeauthoryear{{Hijikawa}, {Kinugawa}, {Yoshida}  \& {Umeda}}{{Hijikawa} et~al.}{2019}]{2019ApJ...882...93H}
{Hijikawa} K.,  {Kinugawa} T.,  {Yoshida} T.,   {Umeda} H.,  2019, \mn@doi [\apj] {10.3847/1538-4357/ab321c}, \href {https://ui.adsabs.harvard.edu/abs/2019ApJ...882...93H} {882, 93}

\bibitem[\protect\citeauthoryear{{Hillman}, {Michaelis}  \& {Perets}}{{Hillman} et~al.}{2025}]{2025arXiv250312586H}
{Hillman} Y.,  {Michaelis} A.,   {Perets} H.~B.,  2025, \mn@doi [arXiv e-prints] {10.48550/arXiv.2503.12586}, \href {https://ui.adsabs.harvard.edu/abs/2025arXiv250312586H} {p. arXiv:2503.12586}

\bibitem[\protect\citeauthoryear{{Hirai} \& {Mandel}}{{Hirai} \& {Mandel}}{2022}]{2022ApJ...937L..42H}
{Hirai} R.,  {Mandel} I.,  2022, \mn@doi [\apjl] {10.3847/2041-8213/ac9519}, \href {https://ui.adsabs.harvard.edu/abs/2022ApJ...937L..42H} {937, L42}

\bibitem[\protect\citeauthoryear{{Hirai}, {Podsiadlowski}  \& {Yamada}}{{Hirai} et~al.}{2018}]{Hirai18}
{Hirai} R.,  {Podsiadlowski} P.,   {Yamada} S.,  2018, \mn@doi [\apj] {10.3847/1538-4357/aad6a0}, \href {https://ui.adsabs.harvard.edu/abs/2018ApJ...864..119H} {864, 119}

\bibitem[\protect\citeauthoryear{{Hiramatsu} et~al.,}{{Hiramatsu} et~al.}{2021}]{2021NatAs...5..903H}
{Hiramatsu} D.,  et~al., 2021, \mn@doi [Nature Astronomy] {10.1038/s41550-021-01384-2}, \href {https://ui.adsabs.harvard.edu/abs/2021NatAs...5..903H} {5, 903}

\bibitem[\protect\citeauthoryear{{Hobbs}, {Lorimer}, {Lyne}  \& {Kramer}}{{Hobbs} et~al.}{2005}]{2005MNRAS.360..974H}
{Hobbs} G.,  {Lorimer} D.~R.,  {Lyne} A.~G.,   {Kramer} M.,  2005, \mn@doi [\mnras] {10.1111/j.1365-2966.2005.09087.x}, \href {https://ui.adsabs.harvard.edu/abs/2005MNRAS.360..974H} {360, 974}

\bibitem[\protect\citeauthoryear{{Hosseinzadeh}, {McCully}, {Zabludoff}, {Arcavi}, {French}, {Howell}, {Berger}  \& {Hiramatsu}}{{Hosseinzadeh} et~al.}{2019}]{2019ApJ...871L...9H}
{Hosseinzadeh} G.,  {McCully} C.,  {Zabludoff} A.~I.,  {Arcavi} I.,  {French} K.~D.,  {Howell} D.~A.,  {Berger} E.,   {Hiramatsu} D.,  2019, \mn@doi [\apjl] {10.3847/2041-8213/aafc61}, \href {https://ui.adsabs.harvard.edu/abs/2019ApJ...871L...9H} {871, L9}

\bibitem[\protect\citeauthoryear{{Hurley}, {Pols}  \& {Tout}}{{Hurley} et~al.}{2000}]{2000MNRAS.315..543H}
{Hurley} J.~R.,  {Pols} O.~R.,   {Tout} C.~A.,  2000, \mn@doi [\mnras] {10.1046/j.1365-8711.2000.03426.x}, \href {https://ui.adsabs.harvard.edu/abs/2000MNRAS.315..543H} {315, 543}

\bibitem[\protect\citeauthoryear{{Hurley}, {Tout}  \& {Pols}}{{Hurley} et~al.}{2002}]{2002MNRAS.329..897H}
{Hurley} J.~R.,  {Tout} C.~A.,   {Pols} O.~R.,  2002, \mn@doi [\mnras] {10.1046/j.1365-8711.2002.05038.x}, \href {https://ui.adsabs.harvard.edu/abs/2002MNRAS.329..897H} {329, 897}

\bibitem[\protect\citeauthoryear{{Jiang}, {Tauris}, {Chen}  \& {Fuller}}{{Jiang} et~al.}{2021}]{2021ApJ...920L..36J}
{Jiang} L.,  {Tauris} T.~M.,  {Chen} W.-C.,   {Fuller} J.,  2021, \mn@doi [\apjl] {10.3847/2041-8213/ac2cc9}, \href {https://ui.adsabs.harvard.edu/abs/2021ApJ...920L..36J} {920, L36}

\bibitem[\protect\citeauthoryear{{Kinugawa} \& {Asano}}{{Kinugawa} \& {Asano}}{2017}]{Kinugawa2017ApJ...849L..29K}
{Kinugawa} T.,  {Asano} K.,  2017, \mn@doi [\apjl] {10.3847/2041-8213/aa95bb}, \href {https://ui.adsabs.harvard.edu/abs/2017ApJ...849L..29K} {849, L29}

\bibitem[\protect\citeauthoryear{{Kinugawa}, {Inayoshi}, {Hotokezaka}, {Nakauchi}  \& {Nakamura}}{{Kinugawa} et~al.}{2014}]{2014MNRAS.442.2963K}
{Kinugawa} T.,  {Inayoshi} K.,  {Hotokezaka} K.,  {Nakauchi} D.,   {Nakamura} T.,  2014, \mn@doi [\mnras] {10.1093/mnras/stu1022}, \href {https://ui.adsabs.harvard.edu/abs/2014MNRAS.442.2963K} {442, 2963}

\bibitem[\protect\citeauthoryear{{Kinugawa}, {Miyamoto}, {Kanda}  \& {Nakamura}}{{Kinugawa} et~al.}{2016}]{2016MNRAS.456.1093K}
{Kinugawa} T.,  {Miyamoto} A.,  {Kanda} N.,   {Nakamura} T.,  2016, \mn@doi [\mnras] {10.1093/mnras/stv2624}, \href {https://ui.adsabs.harvard.edu/abs/2016MNRAS.456.1093K} {456, 1093}

\bibitem[\protect\citeauthoryear{{Kinugawa}, {Horiuchi}, {Takiwaki}  \& {Kotake}}{{Kinugawa} et~al.}{2024}]{2024MNRAS.532.3926K}
{Kinugawa} T.,  {Horiuchi} S.,  {Takiwaki} T.,   {Kotake} K.,  2024, \mn@doi [\mnras] {10.1093/mnras/stae1681}, \href {https://ui.adsabs.harvard.edu/abs/2024MNRAS.532.3926K} {532, 3926}

\bibitem[\protect\citeauthoryear{{Kobulnicky} \& {Fryer}}{{Kobulnicky} \& {Fryer}}{2007}]{2007ApJ...670..747K}
{Kobulnicky} H.~A.,  {Fryer} C.~L.,  2007, \mn@doi [\apj] {10.1086/522073}, \href {https://ui.adsabs.harvard.edu/abs/2007ApJ...670..747K} {670, 747}

\bibitem[\protect\citeauthoryear{{Kobulnicky} et~al.,}{{Kobulnicky} et~al.}{2012}]{2012ApJ...756...50K}
{Kobulnicky} H.~A.,  et~al., 2012, \mn@doi [\apj] {10.1088/0004-637X/756/1/50}, \href {https://ui.adsabs.harvard.edu/abs/2012ApJ...756...50K} {756, 50}

\bibitem[\protect\citeauthoryear{{Laplace}, {G{\"o}tberg}, {de Mink}, {Justham}  \& {Farmer}}{{Laplace} et~al.}{2020}]{2020A&A...637A...6L}
{Laplace} E.,  {G{\"o}tberg} Y.,  {de Mink} S.~E.,  {Justham} S.,   {Farmer} R.,  2020, \mn@doi [\aap] {10.1051/0004-6361/201937300}, \href {https://ui.adsabs.harvard.edu/abs/2020A&A...637A...6L} {637, A6}

\bibitem[\protect\citeauthoryear{{Li} et~al.,}{{Li} et~al.}{2011}]{2011MNRAS.412.1441L}
{Li} W.,  et~al., 2011, \mn@doi [\mnras] {10.1111/j.1365-2966.2011.18160.x}, \href {https://ui.adsabs.harvard.edu/abs/2011MNRAS.412.1441L} {412, 1441}

\bibitem[\protect\citeauthoryear{{Lyman}, {Bersier}, {James}, {Mazzali}, {Eldridge}, {Fraser}  \& {Pian}}{{Lyman} et~al.}{2016}]{Lyman16}
{Lyman} J.~D.,  {Bersier} D.,  {James} P.~A.,  {Mazzali} P.~A.,  {Eldridge} J.~J.,  {Fraser} M.,   {Pian} E.,  2016, \mn@doi [\mnras] {10.1093/mnras/stv2983}, \href {https://ui.adsabs.harvard.edu/abs/2016MNRAS.457..328L} {457, 328}

\bibitem[\protect\citeauthoryear{{Ma} et~al.,}{{Ma} et~al.}{2025}]{Ma25}
{Ma} X.,  et~al., 2025, \mn@doi [arXiv e-prints] {10.48550/arXiv.2504.04393}, \href {https://ui.adsabs.harvard.edu/abs/2025arXiv250404393M} {p. arXiv:2504.04393}

\bibitem[\protect\citeauthoryear{{Maeda} \& {Moriya}}{{Maeda} \& {Moriya}}{2022}]{2022ApJ...927...25M}
{Maeda} K.,  {Moriya} T.~J.,  2022, \mn@doi [\apj] {10.3847/1538-4357/ac4672}, \href {https://ui.adsabs.harvard.edu/abs/2022ApJ...927...25M} {927, 25}

\bibitem[\protect\citeauthoryear{{Maund}, {Pastorello}, {Mattila}, {Itagaki}  \& {Boles}}{{Maund} et~al.}{2016}]{2016ApJ...833..128M}
{Maund} J.~R.,  {Pastorello} A.,  {Mattila} S.,  {Itagaki} K.,   {Boles} T.,  2016, \mn@doi [\apj] {10.3847/1538-4357/833/2/128}, \href {https://ui.adsabs.harvard.edu/abs/2016ApJ...833..128M} {833, 128}

\bibitem[\protect\citeauthoryear{{Metzger}}{{Metzger}}{2022}]{2022ApJ...932...84M}
{Metzger} B.~D.,  2022, \mn@doi [\apj] {10.3847/1538-4357/ac6d59}, \href {https://ui.adsabs.harvard.edu/abs/2022ApJ...932...84M} {932, 84}

\bibitem[\protect\citeauthoryear{{Miyaji}, {Nomoto}, {Yokoi}  \& {Sugimoto}}{{Miyaji} et~al.}{1980}]{1980PASJ...32..303M}
{Miyaji} S.,  {Nomoto} K.,  {Yokoi} K.,   {Sugimoto} D.,  1980, \pasj, \href {https://ui.adsabs.harvard.edu/abs/1980PASJ...32..303M} {32, 303}

\bibitem[\protect\citeauthoryear{{Nagarajan} \& {El-Badry}}{{Nagarajan} \& {El-Badry}}{2024}]{Nagarajan24}
{Nagarajan} P.,  {El-Badry} K.,  2024, \mn@doi [arXiv e-prints] {10.48550/arXiv.2411.16847}, \href {https://ui.adsabs.harvard.edu/abs/2024arXiv241116847N} {p. arXiv:2411.16847}

\bibitem[\protect\citeauthoryear{{Neustadt}, {Kochanek}, {Stanek}, {Basinger}, {Jayasinghe}, {Garling}, {Adams}  \& {Gerke}}{{Neustadt} et~al.}{2021}]{Neustadt21}
{Neustadt} J.~M.~M.,  {Kochanek} C.~S.,  {Stanek} K.~Z.,  {Basinger} C.,  {Jayasinghe} T.,  {Garling} C.~T.,  {Adams} S.~M.,   {Gerke} J.,  2021, \mn@doi [\mnras] {10.1093/mnras/stab2605}, \href {https://ui.adsabs.harvard.edu/abs/2021MNRAS.508..516N} {508, 516}

\bibitem[\protect\citeauthoryear{{Ogata}, {Hirai}  \& {Hijikawa}}{{Ogata} et~al.}{2021}]{2021MNRAS.505.2485O}
{Ogata} M.,  {Hirai} R.,   {Hijikawa} K.,  2021, \mn@doi [\mnras] {10.1093/mnras/stab1439}, \href {https://ui.adsabs.harvard.edu/abs/2021MNRAS.505.2485O} {505, 2485}

\bibitem[\protect\citeauthoryear{{Paczy{\'n}ski}}{{Paczy{\'n}ski}}{1971}]{Paczynski1971}
{Paczy{\'n}ski} B.,  1971, \actaa, \href {https://ui.adsabs.harvard.edu/abs/1971AcA....21....1P} {21, 1}

\bibitem[\protect\citeauthoryear{{Pastorello} et~al.,}{{Pastorello} et~al.}{2007}]{2007Natur.447..829P}
{Pastorello} A.,  et~al., 2007, \mn@doi [\nat] {10.1038/nature05825}, \href {https://ui.adsabs.harvard.edu/abs/2007Natur.447..829P} {447, 829}

\bibitem[\protect\citeauthoryear{{Pastorello} et~al.,}{{Pastorello} et~al.}{2008}]{2008MNRAS.389..113P}
{Pastorello} A.,  et~al., 2008, \mn@doi [\mnras] {10.1111/j.1365-2966.2008.13602.x}, \href {https://ui.adsabs.harvard.edu/abs/2008MNRAS.389..113P} {389, 113}

\bibitem[\protect\citeauthoryear{{Pellegrino} et~al.,}{{Pellegrino} et~al.}{2022}]{2022ApJ...938...73P}
{Pellegrino} C.,  et~al., 2022, \mn@doi [\apj] {10.3847/1538-4357/ac8ff6}, \href {https://ui.adsabs.harvard.edu/abs/2022ApJ...938...73P} {938, 73}

\bibitem[\protect\citeauthoryear{{Perley} et~al.,}{{Perley} et~al.}{2020}]{2020ApJ...904...35P}
{Perley} D.~A.,  et~al., 2020, \mn@doi [\apj] {10.3847/1538-4357/abbd98}, \href {https://ui.adsabs.harvard.edu/abs/2020ApJ...904...35P} {904, 35}

\bibitem[\protect\citeauthoryear{{Pfahl}, {Rappaport}  \& {Podsiadlowski}}{{Pfahl} et~al.}{2002}]{2002ApJ...571L..37P}
{Pfahl} E.,  {Rappaport} S.,   {Podsiadlowski} P.,  2002, \mn@doi [\apjl] {10.1086/341197}, \href {https://ui.adsabs.harvard.edu/abs/2002ApJ...571L..37P} {571, L37}

\bibitem[\protect\citeauthoryear{{Podsiadlowski}, {Joss}  \& {Hsu}}{{Podsiadlowski} et~al.}{1992}]{1992ApJ...391..246P}
{Podsiadlowski} P.,  {Joss} P.~C.,   {Hsu} J.~J.~L.,  1992, \mn@doi [\apj] {10.1086/171341}, \href {https://ui.adsabs.harvard.edu/abs/1992ApJ...391..246P} {391, 246}

\bibitem[\protect\citeauthoryear{{Podsiadlowski}, {Langer}, {Poelarends}, {Rappaport}, {Heger}  \& {Pfahl}}{{Podsiadlowski} et~al.}{2004}]{2004ApJ...612.1044P}
{Podsiadlowski} P.,  {Langer} N.,  {Poelarends} A.~J.~T.,  {Rappaport} S.,  {Heger} A.,   {Pfahl} E.,  2004, \mn@doi [\apj] {10.1086/421713}, \href {https://ui.adsabs.harvard.edu/abs/2004ApJ...612.1044P} {612, 1044}

\bibitem[\protect\citeauthoryear{{Quataert} \& {Shiode}}{{Quataert} \& {Shiode}}{2012}]{Quataert2012}
{Quataert} E.,  {Shiode} J.,  2012, \mn@doi [\mnras] {10.1111/j.1745-3933.2012.01264.x}, \href {https://ui.adsabs.harvard.edu/abs/2012MNRAS.423L..92Q} {423, L92}

\bibitem[\protect\citeauthoryear{{Quataert}, {Lecoanet}  \& {Coughlin}}{{Quataert} et~al.}{2019}]{Quataert19}
{Quataert} E.,  {Lecoanet} D.,   {Coughlin} E.~R.,  2019, \mn@doi [\mnras] {10.1093/mnrasl/slz031}, \href {https://ui.adsabs.harvard.edu/abs/2019MNRAS.485L..83Q} {485, L83}

\bibitem[\protect\citeauthoryear{{Rodr{\'\i}guez}, {Maoz}  \& {Nakar}}{{Rodr{\'\i}guez} et~al.}{2023}]{2023ApJ...955...71R}
{Rodr{\'\i}guez} {\'O}.,  {Maoz} D.,   {Nakar} E.,  2023, \mn@doi [\apj] {10.3847/1538-4357/ace2bd}, \href {https://ui.adsabs.harvard.edu/abs/2023ApJ...955...71R} {955, 71}

\bibitem[\protect\citeauthoryear{{Salpeter}}{{Salpeter}}{1955}]{1955ApJ...121..161S}
{Salpeter} E.~E.,  1955, \mn@doi [\apj] {10.1086/145971}, \href {https://ui.adsabs.harvard.edu/abs/1955ApJ...121..161S} {121, 161}

\bibitem[\protect\citeauthoryear{{Sana} et~al.,}{{Sana} et~al.}{2013}]{2013A&A...550A.107S}
{Sana} H.,  et~al., 2013, \mn@doi [\aap] {10.1051/0004-6361/201219621}, \href {https://ui.adsabs.harvard.edu/abs/2013A&A...550A.107S} {550, A107}

\bibitem[\protect\citeauthoryear{{Sanders} et~al.,}{{Sanders} et~al.}{2013}]{2013ApJ...769...39S}
{Sanders} N.~E.,  et~al., 2013, \mn@doi [\apj] {10.1088/0004-637X/769/1/39}, \href {https://ui.adsabs.harvard.edu/abs/2013ApJ...769...39S} {769, 39}

\bibitem[\protect\citeauthoryear{{Schlegel}}{{Schlegel}}{1990}]{1990MNRAS.244..269S}
{Schlegel} E.~M.,  1990, \mnras, \href {https://ui.adsabs.harvard.edu/abs/1990MNRAS.244..269S} {244, 269}

\bibitem[\protect\citeauthoryear{{Schr{\o}der}, {MacLeod}, {Loeb}, {Vigna-G{\'o}mez}  \& {Mandel}}{{Schr{\o}der} et~al.}{2020}]{Schroder20}
{Schr{\o}der} S.~L.,  {MacLeod} M.,  {Loeb} A.,  {Vigna-G{\'o}mez} A.,   {Mandel} I.,  2020, \mn@doi [\apj] {10.3847/1538-4357/ab7014}, \href {https://ui.adsabs.harvard.edu/abs/2020ApJ...892...13S} {892, 13}

\bibitem[\protect\citeauthoryear{{Schulze} et~al.,}{{Schulze} et~al.}{2021}]{2021ApJS..255...29S}
{Schulze} S.,  et~al., 2021, \mn@doi [\apjs] {10.3847/1538-4365/abff5e}, \href {https://ui.adsabs.harvard.edu/abs/2021ApJS..255...29S} {255, 29}

\bibitem[\protect\citeauthoryear{{Schulze} et~al.,}{{Schulze} et~al.}{2024}]{2024arXiv240902054S}
{Schulze} S.,  et~al., 2024, \mn@doi [arXiv e-prints] {10.48550/arXiv.2409.02054}, \href {https://ui.adsabs.harvard.edu/abs/2024arXiv240902054S} {p. arXiv:2409.02054}

\bibitem[\protect\citeauthoryear{{Shenar} et~al.,}{{Shenar} et~al.}{2022}]{2022A&A...665A.148S}
{Shenar} T.,  et~al., 2022, \mn@doi [\aap] {10.1051/0004-6361/202244245}, \href {https://ui.adsabs.harvard.edu/abs/2022A&A...665A.148S} {665, A148}

\bibitem[\protect\citeauthoryear{{Shiode} \& {Quataert}}{{Shiode} \& {Quataert}}{2014}]{Shiode2014}
{Shiode} J.~H.,  {Quataert} E.,  2014, \mn@doi [\apj] {10.1088/0004-637X/780/1/96}, \href {https://ui.adsabs.harvard.edu/abs/2014ApJ...780...96S} {780, 96}

\bibitem[\protect\citeauthoryear{{Shivvers} et~al.,}{{Shivvers} et~al.}{2017}]{2017MNRAS.471.4381S}
{Shivvers} I.,  et~al., 2017, \mn@doi [\mnras] {10.1093/mnras/stx1885}, \href {https://ui.adsabs.harvard.edu/abs/2017MNRAS.471.4381S} {471, 4381}

\bibitem[\protect\citeauthoryear{{Smartt}}{{Smartt}}{2015}]{2015PASA...32...16S}
{Smartt} S.~J.,  2015, \mn@doi [\pasa] {10.1017/pasa.2015.17}, \href {https://ui.adsabs.harvard.edu/abs/2015PASA...32...16S} {32, e016}

\bibitem[\protect\citeauthoryear{{Smith} \& {Arnett}}{{Smith} \& {Arnett}}{2014}]{Smith2014turb}
{Smith} N.,  {Arnett} W.~D.,  2014, \mn@doi [\apj] {10.1088/0004-637X/785/2/82}, \href {https://ui.adsabs.harvard.edu/abs/2014ApJ...785...82S} {785, 82}

\bibitem[\protect\citeauthoryear{{Smith}, {Li}, {Filippenko}  \& {Chornock}}{{Smith} et~al.}{2011}]{2011MNRAS.412.1522S}
{Smith} N.,  {Li} W.,  {Filippenko} A.~V.,   {Chornock} R.,  2011, \mn@doi [\mnras] {10.1111/j.1365-2966.2011.17229.x}, \href {https://ui.adsabs.harvard.edu/abs/2011MNRAS.412.1522S} {412, 1522}

\bibitem[\protect\citeauthoryear{{Smith} et~al.,}{{Smith} et~al.}{2020}]{2020PASP..132h5002S}
{Smith} K.~W.,  et~al., 2020, \mn@doi [\pasp] {10.1088/1538-3873/ab936e}, \href {https://ui.adsabs.harvard.edu/abs/2020PASP..132h5002S} {132, 085002}

\bibitem[\protect\citeauthoryear{{Soker}}{{Soker}}{2024}]{Soker24}
{Soker} N.,  2024, \mn@doi [The Open Journal of Astrophysics] {10.33232/001c.117147}, \href {https://ui.adsabs.harvard.edu/abs/2024OJAp....7E..31S} {7, 31}

\bibitem[\protect\citeauthoryear{{Sun}, {Maund}, {Hirai}, {Crowther}  \& {Podsiadlowski}}{{Sun} et~al.}{2020}]{Sun20}
{Sun} N.-C.,  {Maund} J.~R.,  {Hirai} R.,  {Crowther} P.~A.,   {Podsiadlowski} P.,  2020, \mn@doi [\mnras] {10.1093/mnras/stz3431}, \href {https://ui.adsabs.harvard.edu/abs/2020MNRAS.491.6000S} {491, 6000}

\bibitem[\protect\citeauthoryear{{Taddia} et~al.,}{{Taddia} et~al.}{2015}]{2015A&A...580A.131T}
{Taddia} F.,  et~al., 2015, \mn@doi [\aap] {10.1051/0004-6361/201525989}, \href {https://ui.adsabs.harvard.edu/abs/2015A&A...580A.131T} {580, A131}

\bibitem[\protect\citeauthoryear{{Taddia} et~al.,}{{Taddia} et~al.}{2018}]{2018A&A...609A.136T}
{Taddia} F.,  et~al., 2018, \mn@doi [\aap] {10.1051/0004-6361/201730844}, \href {https://ui.adsabs.harvard.edu/abs/2018A&A...609A.136T} {609, A136}

\bibitem[\protect\citeauthoryear{{Takahashi}, {Yoshida}  \& {Umeda}}{{Takahashi} et~al.}{2013}]{2013ApJ...771...28T}
{Takahashi} K.,  {Yoshida} T.,   {Umeda} H.,  2013, \mn@doi [\apj] {10.1088/0004-637X/771/1/28}, \href {https://ui.adsabs.harvard.edu/abs/2013ApJ...771...28T} {771, 28}

\bibitem[\protect\citeauthoryear{{Tauris}, {Langer}, {Moriya}, {Podsiadlowski}, {Yoon}  \& {Blinnikov}}{{Tauris} et~al.}{2013}]{2013ApJ...778L..23T}
{Tauris} T.~M.,  {Langer} N.,  {Moriya} T.~J.,  {Podsiadlowski} P.,  {Yoon} S.~C.,   {Blinnikov} S.~I.,  2013, \mn@doi [\apjl] {10.1088/2041-8205/778/2/L23}, \href {https://ui.adsabs.harvard.edu/abs/2013ApJ...778L..23T} {778, L23}

\bibitem[\protect\citeauthoryear{{Tauris}, {Langer}  \& {Podsiadlowski}}{{Tauris} et~al.}{2015}]{2015MNRAS.451.2123T}
{Tauris} T.~M.,  {Langer} N.,   {Podsiadlowski} P.,  2015, \mn@doi [\mnras] {10.1093/mnras/stv990}, \href {https://ui.adsabs.harvard.edu/abs/2015MNRAS.451.2123T} {451, 2123}

\bibitem[\protect\citeauthoryear{{Tian} et~al.,}{{Tian} et~al.}{2018}]{Tian2018}
{Tian} Z.-J.,  et~al., 2018, \mn@doi [Research in Astronomy and Astrophysics] {10.1088/1674-4527/18/5/52}, \href {https://ui.adsabs.harvard.edu/abs/2018RAA....18...52T} {18, 052}

\bibitem[\protect\citeauthoryear{{Tominaga} et~al.,}{{Tominaga} et~al.}{2008}]{2008ApJ...687.1208T}
{Tominaga} N.,  et~al., 2008, \mn@doi [\apj] {10.1086/591782}, \href {https://ui.adsabs.harvard.edu/abs/2008ApJ...687.1208T} {687, 1208}

\bibitem[\protect\citeauthoryear{{Tsuna}, {Wu}, {Fuller}, {Dong}  \& {Piro}}{{Tsuna} et~al.}{2024}]{2024OJAp....7E..82T}
{Tsuna} D.,  {Wu} S.~C.,  {Fuller} J.,  {Dong} Y.,   {Piro} A.~L.,  2024, \mn@doi [The Open Journal of Astrophysics] {10.33232/001c.123897}, \href {https://ui.adsabs.harvard.edu/abs/2024OJAp....7E..82T} {7, 82}

\bibitem[\protect\citeauthoryear{{Tuna} \& {Metzger}}{{Tuna} \& {Metzger}}{2023}]{2023ApJ...955..125T}
{Tuna} S.,  {Metzger} B.~D.,  2023, \mn@doi [\apj] {10.3847/1538-4357/acef17}, \href {https://ui.adsabs.harvard.edu/abs/2023ApJ...955..125T} {955, 125}

\bibitem[\protect\citeauthoryear{{Vink}, {de Koter}  \& {Lamers}}{{Vink} et~al.}{2001}]{Vink2001A&A...369..574V}
{Vink} J.~S.,  {de Koter} A.,   {Lamers} H.~J.~G.~L.~M.,  2001, \mn@doi [\aap] {10.1051/0004-6361:20010127}, \href {https://ui.adsabs.harvard.edu/abs/2001A&A...369..574V} {369, 574}

\bibitem[\protect\citeauthoryear{{Warwick} et~al.,}{{Warwick} et~al.}{2025}]{2025MNRAS.536.3588W}
{Warwick} B.,  et~al., 2025, \mn@doi [\mnras] {10.1093/mnras/stae2784}, \href {https://ui.adsabs.harvard.edu/abs/2025MNRAS.536.3588W} {536, 3588}

\bibitem[\protect\citeauthoryear{{Webbink}}{{Webbink}}{1984}]{1984ApJ...277..355W}
{Webbink} R.~F.,  1984, \mn@doi [\apj] {10.1086/161701}, \href {https://ui.adsabs.harvard.edu/abs/1984ApJ...277..355W} {277, 355}

\bibitem[\protect\citeauthoryear{{Wei}, {Schneider}, {Podsiadlowski}, {Laplace}, {R{\"o}pke}  \& {Vetter}}{{Wei} et~al.}{2024}]{2024A&A...688A..87W}
{Wei} D.,  {Schneider} F. R.~N.,  {Podsiadlowski} P.,  {Laplace} E.,  {R{\"o}pke} F.~K.,   {Vetter} M.,  2024, \mn@doi [\aap] {10.1051/0004-6361/202348560}, \href {https://ui.adsabs.harvard.edu/abs/2024A&A...688A..87W} {688, A87}

\bibitem[\protect\citeauthoryear{{Wheeler}, {Lecar}  \& {McKee}}{{Wheeler} et~al.}{1975}]{Wheeler75}
{Wheeler} J.~C.,  {Lecar} M.,   {McKee} C.~F.,  1975, \mn@doi [\apj] {10.1086/153771}, \href {https://ui.adsabs.harvard.edu/abs/1975ApJ...200..145W} {200, 145}

\bibitem[\protect\citeauthoryear{{Willcox}, {Mandel}, {Thrane}, {Deller}, {Stevenson}  \& {Vigna-G{\'o}mez}}{{Willcox} et~al.}{2021}]{2021ApJ...920L..37W}
{Willcox} R.,  {Mandel} I.,  {Thrane} E.,  {Deller} A.,  {Stevenson} S.,   {Vigna-G{\'o}mez} A.,  2021, \mn@doi [\apjl] {10.3847/2041-8213/ac2cc8}, \href {https://ui.adsabs.harvard.edu/abs/2021ApJ...920L..37W} {920, L37}

\bibitem[\protect\citeauthoryear{{Woosley}}{{Woosley}}{2019}]{Woosley2019}
{Woosley} S.~E.,  2019, \mn@doi [\apj] {10.3847/1538-4357/ab1b41}, \href {https://ui.adsabs.harvard.edu/abs/2019ApJ...878...49W} {878, 49}

\bibitem[\protect\citeauthoryear{{Woosley}, {Heger}  \& {Weaver}}{{Woosley} et~al.}{2002}]{Woosley2002}
{Woosley} S.~E.,  {Heger} A.,   {Weaver} T.~A.,  2002, \mn@doi [Reviews of Modern Physics] {10.1103/RevModPhys.74.1015}, \href {https://ui.adsabs.harvard.edu/abs/2002RvMP...74.1015W} {74, 1015}

\bibitem[\protect\citeauthoryear{{Woosley}, {Blinnikov}  \& {Heger}}{{Woosley} et~al.}{2007}]{Woosley2007}
{Woosley} S.~E.,  {Blinnikov} S.,   {Heger} A.,  2007, \mn@doi [\nat] {10.1038/nature06333}, \href {https://ui.adsabs.harvard.edu/abs/2007Natur.450..390W} {450, 390}

\bibitem[\protect\citeauthoryear{{Wu} \& {Fuller}}{{Wu} \& {Fuller}}{2022}]{wu2022extreme}
{Wu} S.~C.,  {Fuller} J.,  2022, \mn@doi [\apjl] {10.3847/2041-8213/ac9b3d}, \href {https://ui.adsabs.harvard.edu/abs/2022ApJ...940L..27W} {940, L27}

\bibitem[\protect\citeauthoryear{{Yoon}, {Woosley}  \& {Langer}}{{Yoon} et~al.}{2010}]{2010ApJ...725..940Y}
{Yoon} S.~C.,  {Woosley} S.~E.,   {Langer} N.,  2010, \mn@doi [\apj] {10.1088/0004-637X/725/1/940}, \href {https://ui.adsabs.harvard.edu/abs/2010ApJ...725..940Y} {725, 940}

\bibitem[\protect\citeauthoryear{{Zapartas} et~al.,}{{Zapartas} et~al.}{2017a}]{2017A&A...601A..29Z}
{Zapartas} E.,  et~al., 2017a, \mn@doi [\aap] {10.1051/0004-6361/201629685}, \href {https://ui.adsabs.harvard.edu/abs/2017A&A...601A..29Z} {601, A29}

\bibitem[\protect\citeauthoryear{{Zapartas} et~al.,}{{Zapartas} et~al.}{2017b}]{2017ApJ...842..125Z}
{Zapartas} E.,  et~al., 2017b, \mn@doi [\apj] {10.3847/1538-4357/aa7467}, \href {https://ui.adsabs.harvard.edu/abs/2017ApJ...842..125Z} {842, 125}

\bibitem[\protect\citeauthoryear{{Zhang} \& {Fryer}}{{Zhang} \& {Fryer}}{2001}]{Zhang01}
{Zhang} W.,  {Fryer} C.~L.,  2001, \mn@doi [\apj] {10.1086/319734}, \href {https://ui.adsabs.harvard.edu/abs/2001ApJ...550..357Z} {550, 357}

\bibitem[\protect\citeauthoryear{{de Kool}}{{de Kool}}{1990}]{deKool90}
{de Kool} M.,  1990, \mn@doi [\apj] {10.1086/168974}, \href {https://ui.adsabs.harvard.edu/abs/1990ApJ...358..189D} {358, 189}

\bibitem[\protect\citeauthoryear{{van den Heuvel}}{{van den Heuvel}}{2010}]{2010NewAR..54..140V}
{van den Heuvel} E. P.~J.,  2010, \mn@doi [\nar] {10.1016/j.newar.2010.09.031}, \href {https://ui.adsabs.harvard.edu/abs/2010NewAR..54..140V} {54, 140}

\makeatother
\end{thebibliography}



\appendix 

\section{Extraction Methods for CCSNe}
In this section, we report how we extract the events; CCSNe, SESNe, and merger scenario SNe Ibn. Hereafter, we set Star 1 as the heavier star at ZAMS, and Star 2 as the lighter star at ZAMS. We also define "H star" as the star with hydrogen such as MS and AGB star, which corresponds to the stellar types 1 to 6 in the classification by \citet{2002MNRAS.329..897H}. "He star" refers to the classification 7 to 9 in \citet{2002MNRAS.329..897H}). "WD" refers to the classification10 to 12 in \citet{2002MNRAS.329..897H}).   We also set the conditions to be satisfied are marked with a circle ($\bigcirc$), while the conditions that should not be satisfied are marked with a cross ($\times$).

\subsection{Extraction of CCSNe}\label{sec:app_CCSN}
In order to count the number of binary-origin CCSNe, we extract stars that satisfy the following conditions:
\begin{itemize}
    \item[$\bigcirc$] The system includes a NS or BH.
    \item[$\times$] The NS or BH evolved from a WD.
\end{itemize}
We search for cases in which a remnant type appears without being preceded by a WD type at the previous timestep. If such a transition is found, we classify the star as a CCSN. The total number of CCSNe is then computed by subtracting the number of NS/BH remnants originating from WD progenitors from the total number of NS/BH remnants.

This approach distinguishes CCSNe from alternative evolutionary channels such as accretion-induced collapse or WD mergers.
\subsection{Extraction of SESNe}
\label{sec:app_Ibc}
The number of SESNe in our model is same as the number of exploding He star. The exploding He star can either be the lighter star in the binary at ZAMS (Case 1) or the heavier star in the binary at ZAMS (Case 2). Therefore, we counted these cases separately. 

For Case 1, since Star 1 can only be a NS or BH, we extract the exploding He stars based on the following condition;
\begin{itemize}
    \item[$\bigcirc$] A Star 2 that is a He star at a given moment but evolves into a NS or BH in the subsequent time step.
\end{itemize}

For Case 2, we extract the exploding He stars based on the following conditions;
\begin{itemize}
    \item[$\bigcirc$] A Star 1 that is a He star at a given moment but evolves into a NS or BH in the subsequent time step.
    \item[$\times$] At the time when Star 1 becomes a NS or BH, Star 2 is a massless remnant, which correspond to the state 15 in \cite{2002MNRAS.329..897H}.
    \item[$\times$] At a given moment, Star 1 is a He star and Star 2 is an H star, and at the subsequent stage, the mass of the He star (Star 1) increases.
\end{itemize}
When Star 1 is a He star, Star 2 is either an H or a He star. Therefore, mergers between them do not necessarily lead to SNe, and such cases are excluded. This is the reason for the second condition, as in the case of a merger, Star 2 is represented as a massless remnant. 

Additionally, before Star 1 (He star) explodes, it may undergo mass transfer from Star 2 (H star). This happens on rare cases with initial mass ratios close to unity, where Star 2 leaves the main sequence while Star 1 is a He star. During this process, Star 1 remains flagged as a He star in the BSE code, but due to the accretion of a significant amount of H-rich material we expect the progenitor does not explode as SESNe. This is the reason for imposing the last condition.

\subsection{Extraction of SNe Ibn (merger origin)}\label{sec:app_merger}
To number the population of merger scenario SNe Ibn, we extract the binary systems in which a He star merges with a compact object. Specifically, we search for the following condition within the time evolution of each binary component:

\begin{itemize}
    \item[$\bigcirc$] At a given moment, Star 2 is a He star, and Star 1 is either a NS or BH.
    \item[$\bigcirc$] In the subsequent time step, Star 2 evolves into a state with massless remnant.
\end{itemize}
In the merger scenario, we focus on binary systems consisting of a He star and either a NS or a BH. The first step is to extract such systems. During the merger process, the BSE code designates one of the stars, typically the lighter one, as a massless remnant. Therefore, we extract the systems in which this condition is met.

\section{{Ratio Dependence on Mass Range}}\label{sec:app_massrange}
\begin{figure*}
 \centering
 \includegraphics[width=\linewidth]{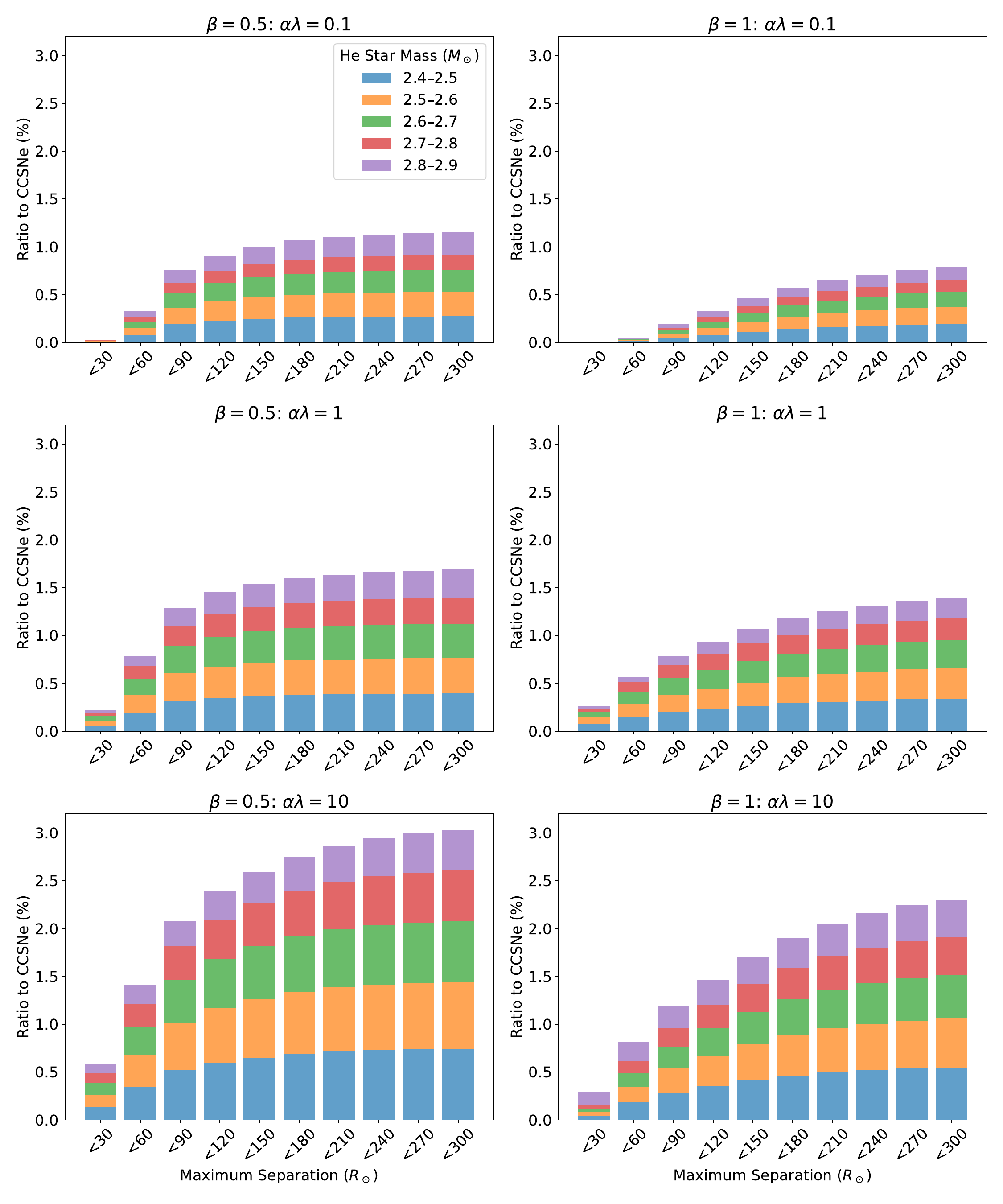}
\caption{{Event rate of SN Ibn relative to CCSNe as in Figure \ref{fig:Ibn_rate}, but for a He star mass range of 2.4--2.9 $M_{\odot}$.}}
\label{fig:Ibn_rate_app1}
\end{figure*}

\begin{figure*}
 \centering
 \includegraphics[width=\linewidth]{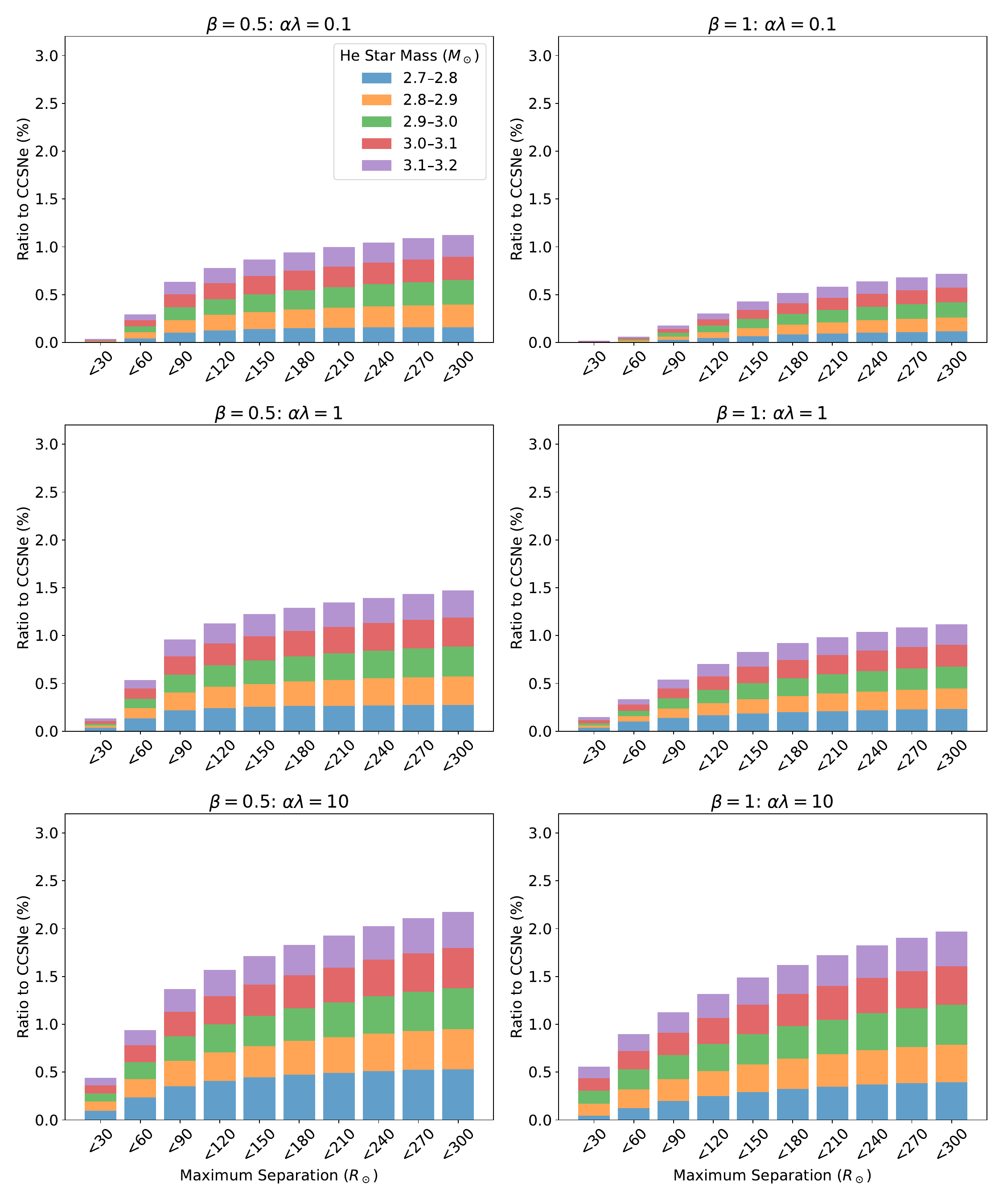}
\caption{{Event rate of SN Ibn relative to CCSNe as in Figure \ref{fig:Ibn_rate}, but for a He star mass range of 2.7--3.2 $M_{\odot}$.}}
\label{fig:Ibn_rate_app2}
\end{figure*}

{The He star mass range $2.5\text{--}3.0\,M_\odot$, which we consider in this work is approximate values. In order to confirm that this range does not affect the results much, we perform the same analysis but for different mass ranges ($2.4\text{--}2.9\,M_\odot$) and ($2.7\text{--}3.2\,M_\odot$). Figures \ref{fig:Ibn_rate_app1} and \ref{fig:Ibn_rate_app2} show the results, which we obtain in the same way as in Figure~\ref{fig:Ibn_rate}. The mass range weakly affects the rates, with slightly higher rates for lower mass thresholds likely reflecting the IMF. However, the variations due to mass range is much weaker than those arising from uncertainties in the binary parameters (e.g. $\alpha\lambda$ and $\beta$).}

\bsp	
\label{lastpage}
\end{document}